\documentclass[lettersize,journal]{IEEEtran}
\usepackage{newtxtext,newtxmath}
\usepackage{booktabs,tabularx}
\usepackage{multirow}
\usepackage{colortbl}
\usepackage{subfigure}
\usepackage{amsmath,amsfonts}
\usepackage{algorithm}
\usepackage{tabularx}
\usepackage[noend]{algpseudocode}
\usepackage{array}
\usepackage[caption=false,font=normalsize,labelfont=sf,textfont=sf]{subfig}
\usepackage{textcomp}
\usepackage{stfloats}
\usepackage{url}
\usepackage{verbatim}
\usepackage{graphicx}
\usepackage{cite}
\usepackage{wasysym}
\usepackage[cmyk]{xcolor}
\begin{document}

\title{Large Language Models Merging for Enhancing the Link Stealing Attack on Graph Neural Networks}

\author{Faqian Guan,~
        Tianqing Zhu*,~\IEEEmembership{Member,~IEEE,}
        Wenhan Chang,~ 
        Wei Ren,~\IEEEmembership{Member,~IEEE,} 
        and~Wanlei Zhou,~\IEEEmembership{Senior Member,~IEEE}~

\IEEEcompsocitemizethanks{\IEEEcompsocthanksitem Faqian Guan and Wenhan Chang are with the China University of Geosciences, Wuhan, China; Tianqing Zhu and Wanlei Zhou are with the City University of Macau, Macau, China. 
\protect\\

\IEEEcompsocthanksitem Tianqing Zhu is the corresponding author. E-mail: tqzhu@cityu.edu.mo}

\thanks{Manuscript received December 9, 2024; revised XX XX, 2024.}}

\markboth{Journal of \LaTeX\ Class Files,~Vol.~14, No.~8, August~2021}%
{Shell \MakeLowercase{\textit{et al.}}: A Sample Article Using IEEEtran.cls for IEEE Journals}

\maketitle

\begin{abstract}
Graph Neural Networks (GNNs), specifically designed to process the graph data, have achieved remarkable success in various applications. Link stealing attacks on graph data pose a significant privacy threat, as attackers aim to extract sensitive relationships between nodes (entities), potentially leading to academic misconduct, fraudulent transactions, or other malicious activities. Previous studies have primarily focused on single datasets and did not explore cross-dataset attacks, let alone attacks that leverage the combined knowledge of multiple attackers. 
However, we find that an attacker can combine the data knowledge of multiple attackers to create a more effective attack model, which can be referred to cross-dataset attacks. Moreover, if knowledge can be extracted with the help of Large Language Models (LLMs), the attack capability will be more significant. 
In this paper, we propose a novel link stealing attack method that takes advantage of cross-dataset and Large Language Models (LLMs). 
The LLM is applied to process datasets with different data structures in cross-dataset attacks. Each attacker fine-tunes the LLM on their specific dataset to generate a tailored attack model.
We then introduce a novel model merging method to integrate the parameters of these attacker-specific models effectively. The result is a merged attack model with superior generalization capabilities, enabling effective attacks not only on the attackers' datasets but also on previously unseen (out-of-domain) datasets. We conducted extensive experiments in four datasets to demonstrate the effectiveness of our method. Additional experiments with three different GNN and LLM architectures further illustrate the generality of our approach. In summary, we present a new link stealing attack method that facilitates collaboration among multiple attackers to develop a powerful, universal attack model that reflects realistic real-world scenarios.

\end{abstract}


\section{Introduction}

\IEEEPARstart{G}{r}aph Neural Networks (GNNs) \cite{DBLP:journals/tnn/ScarselliGTHM09} offer an effective approach for processing data with graph structures. By capturing relational information between nodes in a graph, GNNs integrate these relationships into node representation learning. Alongside research into the applications of GNNs, some researchers also investigate privacy attacks targeting GNNs \cite{DBLP:journals/air/GuanZZC24}. 

Similarly to privacy attacks in the image and text domains, attackers in GNN privacy attacks aim to extract sensitive information. In particular, GNNs are susceptible to a unique type of privacy attack called the link stealing attack \cite{DBLP:conf/uss/HeJ0G021}, where attackers extract the connections between nodes of the training graph data by compromising the target model.
For example, in finance, link stealing attacks can expose associations between certain users in transaction networks, enabling fraudsters to identify potential targets or establish chains of fraudulent transactions and account associations. In social networks, link stealing attacks can help attackers deduce relationships between nodes without direct data access.

Given the practical significance of link stealing attacks, numerous researchers have investigated these attacks on GNNs \cite{DBLP:conf/uss/HeJ0G021,DBLP:conf/icml/ZhangWWYXPY23,DBLP:journals/popets/WuHBHBGZ24,DBLP:journals/corr/abs-2307-13548,DBLP:journals/corr/abs-2406-16963,shandong5}. The primary method in link stealing attacks involves comparing the similarity between the posterior probabilities of different nodes to infer the existence of links \cite{DBLP:conf/uss/HeJ0G021, DBLP:conf/icml/ZhangWWYXPY23,DBLP:journals/popets/WuHBHBGZ24}. Other researchers infer the existence of links by observing changes in the posterior probabilities of nodes after introducing perturbations to the original graph data \cite{DBLP:journals/corr/abs-2307-13548}. 
However, previous studies have not explored the potential of leveraging the knowledge of multiple attackers to collaboratively train a more generalized and effective attack model. Furthermore, due to constraints such as privacy protections, commercial interests, and policy regulations, these attackers cannot share data, which complicates collaborative training efforts. 
Attacking multiple datasets required separate models for each dataset, resulting in substantial training time and resource consumption~\cite{9319526,TIFS3315526}. Developing a universal attack model that can simultaneously target multiple datasets would markedly reduce the need for time, resources, and computational infrastructure. 
In summary, conducting collaborative link stealing attacks by combining multiple attackers would be much more powerful. However, this novel attack presents two key challenges. 
\begin{itemize}
    \item How to implement cross-dataset link stealing attacks.
    \item How to aggregate knowledge from multiple attackers without sharing local data.
\end{itemize}

To address the first challenge, we adopt large language models (LLMs) \cite{DBLP:conf/acl/LiYBZLSLSYWLXBF24} to perform a cross-dataset link stealing attack. LLMs utilize the attention mechanism within the transformer architecture, which is particularly effective at processing variable-length inputs in parallel, allowing LLMs to adapt well to datasets with differing feature dimensions and lengths \cite{MSurvey6}. Additionally, LLMs can capture subtle linguistic nuances, complex semantic structures, and intricate text patterns. These depth of understanding empower LLMs to achieve state-of-the-art performance across a range of tasks \cite{DBLP:conf/iclr/ZengLDWL0YXZXTM23, DBLP:conf/eacl/LinWZLW24, DBLP:conf/wsdm/LiuCS024}. Thus, by introducing LLMs into the development of a link stealing attack model, we aim to address the difficulties associated with cross-dataset attacks while also enhancing overall attack effectiveness.

To address the second challenge, we employ model merging techniques to aggregate knowledge from multiple attackers. Model merging combines the strengths of individual models to produce a single, more robust model. Specifically, each attacker independently trains a model based on their unique dataset. Instead of sharing data, attackers share only their model parameters, which are then merged to create a unified model. This approach allows multiple attackers to collaboratively enhance the effectiveness of link stealing attacks by integrating the strengths of each model, resulting in a universal link stealing attack model without compromising local data privacy.

In this paper, we propose a novel link stealing attack method that integrates large language models (LLMs) and model merging techniques. These methods enable multiple attackers to collaborate and create a more effective attack model capable of performing cross-dataset attacks.
First, we introduce the use of LLMs for link stealing attacks, leveraging their ability to process variable-length data and perform cross-dataset operations. To improve the LLM’s performance in link stealing tasks, we design specific prompts that incorporate both textual features and node posterior probabilities. This allows each attacker to independently fine-tune their LLM-based model using their own dataset and knowledge.
Next, we propose a novel model merging approach to combine the LLM-based attack models developed by multiple attackers and construct a universal link stealing attack model. This approach involves three key steps: LLM parameter dropping, parameter selection, and parameter merging. These steps facilitate the integration of individual attackers' knowledge, resulting in a generalized attack model. Moreover, we demonstrate that the model merging method can effectively target out-of-domain data, datasets for which attackers have no prior knowledge, thus enhancing the model's real-world applicability.
Finally, we validate the effectiveness of our approach through thorough theoretical analysis and extensive experimental evaluation.

In summary, our paper contributes the following.

\begin{itemize}

    \item We propose a novel link stealing attack method that focuses on combining the knowledge of multiple attackers to collaboratively create a more capable and generalized attack model. This approach enables the development of more realistic and effective attacks, better aligned with real-world scenarios.

    \item Our method introduces the use of LLMs for link stealing attacks, leveraging their ability to handle cross-dataset scenarios and achieve improved attack performance.

    \item We propose a novel model merging method to integrate knowledge from multiple attackers, resulting in a universal attack model capable of executing attacks across diverse datasets, including previously unseen (out-of-domain) datasets.

    \item We validate the effectiveness of our method through comprehensive theoretical analysis and experiments across multiple datasets.

\end{itemize}

\section{Preliminary}
\subsection{Notations}
We denote $\mathrm{G}$ as the graph data and $\mathrm{X}$ as the node features in the graph. $\mathrm{N}$ and $\mathrm{T}$ represent the numerical and textual features of nodes, respectively. Nodes in the graph are represented as $u$ and $v$, with $\mathcal{N}(v)$ indicating the neighboring nodes of node $v$. $\mathrm{P}$ denotes the posterior probability of nodes obtained from the target model. $\mathrm{H}$ represents the hidden state of a feature vector, $\mathcal{T}$ refers to the target model, and $\mathcal{A}$ refers to the attack model. $\mathrm{Y}$ and $\hat{\mathrm{Y}}$ denote the ground truth and predicted labels of nodes, respectively. $Link$ and $Unlink$ indicate the presence and absence of a link between nodes. $\theta$ represents the parameters of the pre-trained LLM, while $\delta$ denotes the parameter updates in the LLM model. $\lambda$ is the weight assigned to a model during the merging process, and $d$ is the dropout probability. $\mathrm{W}$ and $\mathrm{B}$ represent the neural network parameters and bias term, respectively, while $\sigma$ is the activation function. Lastly, $\gamma$ is the scaling factor used in the model merging process. The notations used throughout this paper are summarized in Table \ref{Notations}.

\begin{table}[htbp]
  \centering
  \caption{Summary of notations.}
  
  \scalebox{1}{
    \begin{tabular}{c|c}
    \toprule
    \textbf{Notations} & \textbf{ Description} \\
    \midrule
    $\mathrm{G}$ & Graph data \\
    $\mathrm{X(N, T)}$ & Node features \\
    $\mathrm{N}$ & Numerical features of nodes \\
    $\mathrm{T}$ & Textual features of nodes \\
    $u, v$ & Individual nodes in the graph \\
    $\mathcal{N}(v)$ & Neighboring nodes of node $v$ \\
    $\mathrm{P}$ & Posterior probabilities for nodes \\
    $\mathrm{H}$ & Hidden state of a feature vector \\
    $\mathcal{T}$ & Target model \\
    $\mathcal{A}$ & Attack model \\
    $\mathrm{Y}$ & Ground truth labels of nodes \\
    $\hat{\mathrm{Y}}$ & Predicted labels of nodes \\

    $Link$ & Indicates the presence of a link between nodes \\
    $Unlink$ & Indicates the absence of a link between nodes \\
    $\theta$ & Parameters of the pre-trained LLM \\
    $\delta$ & Parameter updates in the LLM model \\
    $\lambda$ & Weight assigned to a model during merging \\
    $d$ & Dropout probability \\
    $\mathrm{W}$ & Neural network parameters \\
    $\mathrm{B}$ & Neural network bias \\
    $\sigma$ & Activation function \\
    $\gamma$ & Scaling factor \\

    \bottomrule
    \end{tabular}%
    }
  \label{Notations}%
\end{table}%

\subsection{Graph Neural Networks}
Graph Neural Networks (GNNs) have emerged as a powerful framework to process data with graph structures. GNNs have demonstrated impressive results in various graph-based applications by learning effective representations of both node features and relational information within the graph. The central mechanism of GNNs is the message passing, or neighborhood aggregation, process, where the information is propagated through the graph structure. Each node updates its representation by aggregating the features of its neighbors and combining them with its own features. The updated representation of a node in the $k$-th layer of the GNN can be expressed as follows:

\begin{equation} \label{gnn_eq}
    \begin{gathered}
        {h}_{v}^{(k)}=\operatorname{UPDATE}^{(k)}\left({h}_{v}^{(k-1)}, {e}_{v}^{(k)}\right) \\
        {e}_{v}^{(k)} = \operatorname {AGG}^{(k-1)}\left(\left\{{h}_{u}^{(k-1)}: \forall u \in \mathcal{N}(v) \cup v\right\}\right)    
    \end{gathered}
\end{equation}
where $\mathcal{N}(v)$ denotes the neighbors of node $v$. ${h}_{v}^{(k)}$ is the representation of node $v$ after the $k$-th layer update, while ${h}_{v}^{(0)}$ refers to the initial input feature (i.e., $x_v$). The function $\operatorname{AGG}(\cdot)$ aggregates information from neighboring nodes, and $\operatorname{UPDATE}(\cdot)$ incorporates this aggregated information into the node's representation.

\subsection{Large Language Models}
Large Language Models (LLMs) \cite{DBLP:conf/acl/LiYBZLSLSYWLXBF24} are sophisticated artificial intelligence models specifically designed for natural language processing tasks. These models use deep learning architectures, often based on transformer models, to understand and generate human-like language by capturing complex linguistic patterns, context, and nuanced meanings in text. They are termed "large" because they are trained on extensive datasets and contain an exceptionally high number of parameters, often reaching billions or even trillions. The more parameters a model has, the more information it can encode, which enhances its performance in a wide range of tasks. However, deploying and managing these large-parameter models in practical applications can be challenging because of their substantial computational requirements.

\subsection{Link Stealing Attack}
\begin{figure}[htp]
\centering
\includegraphics[scale=0.59]{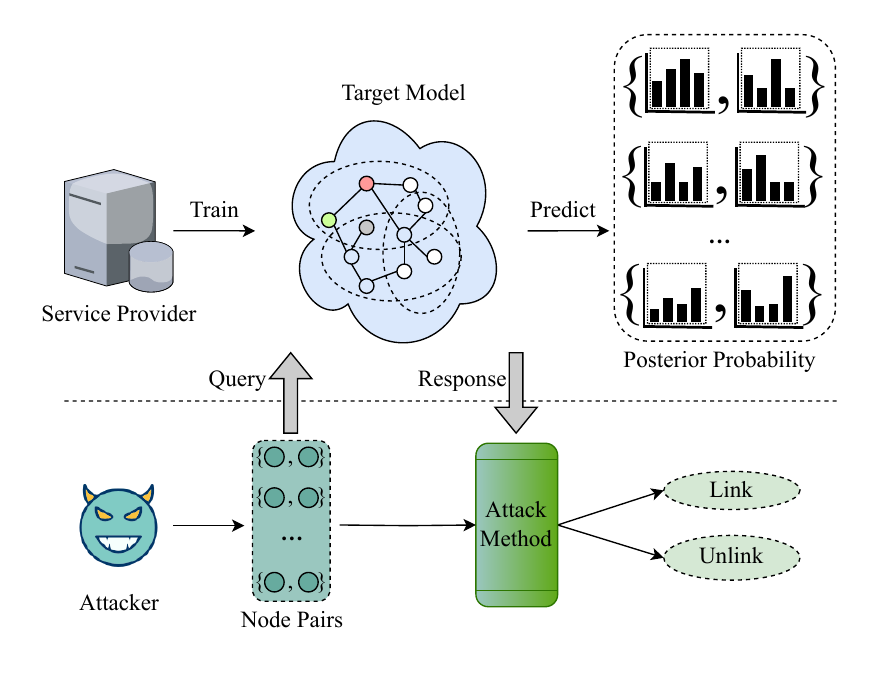}
\caption{Overview of general link stealing attacks. The upper section depicts the service provider offering the target model for user access, while the lower section illustrates attackers executing the attack process.}
\label{link_stealing}
\end{figure}

A link stealing attack is a privacy attack that targets inference of private links between nodes within graph-structured data \cite{DBLP:journals/air/GuanZZC24}. 
In link stealing attacks, attackers aim to infer the existence of links between nodes by leveraging both node knowledge and response information obtained from accessing the target model. Specifically, as illustrated in Fig. \ref{link_stealing}, a service provider trains a model using company data and deploys it for user access, typically to serve commercial purposes. This deployed model becomes the target of link stealing attacks. Attackers first form node pairs by pairing nodes whose connection status they wish to infer. By querying the target model, they then obtain the posterior probabilities of these node pairs. Using the original node information alongside the acquired posterior probabilities, attackers construct a link stealing attack method. This method enables attackers to infer the presence or absence of links between node pairs, posing a significant privacy threat to the service provider.

\subsection{Model Merging}
With the rise of LLMs, researchers have increasingly explored ways to leverage different LLMs trained on separate datasets or tasks through model merging \cite{DBLP:journals/corr/abs-2407-06089}. This approach aims to harness the distinct strengths of each model, creating a more versatile and universal LLM. Model merging combines two or more trained models into a single, unified model that incorporates the benefits of each original model. By merging model parameters, this technique enables the creation of a more capable model without requiring data sharing, thus safeguarding data privacy. One straightforward method for model merging is parameter averaging. Given $n$ models with parameters ${{\theta}_1,{\theta}_2,...,{\theta}_n}$, in the parameter averaging method, the parameters of the merged model can be calculated as:

\begin{equation}
    \begin{gathered}
    {\theta}^{merge}=\frac{1}{n}{\sum}_{i=1}^{n}{\theta}_i
    \end{gathered}
\end{equation}

\section{Problem Definition}

\subsubsection{Attack Goal} In a link stealing attack, attackers use the attack method $\mathcal{A}$ to infer whether there are links between nodes based on response information $R$ obtained by accessing the target model $\mathcal{T}$. Specifically, the attacker aims to determine whether there is a link between nodes $u$ and $v$. First, the attacker inputs the nodes $u$ and $v$ into $\mathcal{T}$ for query. The target model $\mathcal{T}$ then outputs $R$ based on the features of nodes $u$ and $v$. The attacker analyzes the attack method $\mathcal{A}$ and infers whether there is a link between the nodes $u$ and $v$ by combining $\mathrm{R}$ with the features of the nodes $u$ and $v$. This attack can be formally defined as follows.

\begin{equation}
    \begin{gathered}
        \mathcal{A}: \{u, v\}, \{r_u, r_v\} \mapsto \text ( \emph{Link}, \emph{Unlink} )\\
        \mathcal{T}: \{u, v\} \mapsto \{r_u, r_v\}
    \end{gathered}
\end{equation}
In this paper, $r_u$ and $r_v$ represent the posterior probabilities of node predictions for nodes $u$ and $v$ provided by the target model $\mathcal{T}$. \emph{Link} indicates that there is a link between $u$ and $v$, while \emph{Unlink} indicates that there is no link.

\textbf{Target Model.} The target model, $\mathcal{T}$, is also known as the victim model in link stealing attacks. The service provider trains $\mathcal{T}$ on the target dataset and deploys it on the Internet for user access. Attackers then access the deployed model $\mathcal{T}$ and obtain its response information $R$. This information is used to steal node links in the target dataset. In this paper, $\mathcal{T}$ is a node classification model.

\subsubsection{Attack Setting} In this paper, we study link stealing attacks by simulating a real-world scenario. Multiple attackers, or a multinational company with different departments, aim to jointly train a more efficient and robust attack model. However, the attackers do not want to share the data they have collected at a high cost with others. 

\textbf{Attacker Knowledge.} Each attacker or department has its own independently collected dataset, denoted as $G={G_1,G_2,...,G_n}$. These datasets contain links between certain nodes. However, attackers do not have knowledge of the structure or parameters of the target model. They can only interact with the model by sending nodes and obtaining the posterior probabilities of the corresponding nodes. This scenario corresponds to the most threatening black-box setting in privacy attacks \cite{DBLP:journals/air/GuanZZC24}.

In joint training of the attack model, data sharing among attackers is avoided due to the high cost of data collection. Instead, attackers share only their trained model parameters for collaborative training. Specifically, each attacker uses their own dataset and the derived posterior probabilities to train a link stealing attack model. By applying a suitable model merging method, the strengths of each individual model are integrated, achieving the objective of joint training without the need for data exchange.

\begin{figure*}[t]
\centering
\includegraphics[scale=0.73]{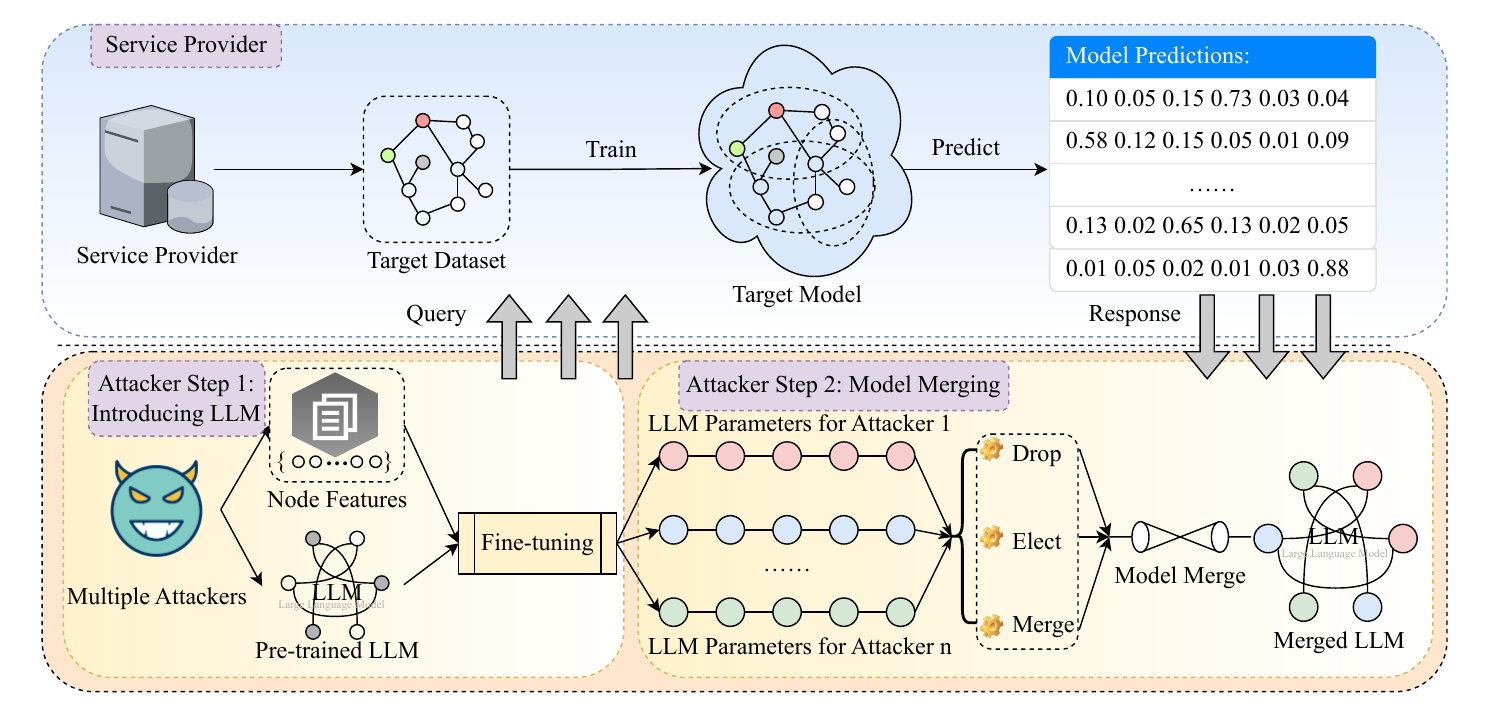}
\caption{Overview of the proposed link stealing attack method. The upper section illustrates the service provider offering the target model for user access. The lower section details the attackers’ process of creating attack models using our proposed method. In Attacker Step 1, Attackers introduce LLMs and leverage their respective node features to fine-tune and develop multiple LLM-based link stealing attack models. In Attacker Step 2, the attackers, through model merging techniques, combine the strengths of these individual models to create a more robust merged LLM model for executing link stealing attacks.}
\label{Overview}
\end{figure*}

\section{Methodology}

\subsection{Attack Overview}
In this paper, we study link stealing attacks, where attackers target a deployed model to infer links from its training dataset. An overview of the proposed link stealing attack is illustrated in Fig. \ref{Overview}. The service provider invests substantial resources to collect the target dataset and train the target model, which is then deployed online for user access to fulfill commercial or profitable objectives.

In our proposed link stealing attack method, we focus on two key steps: Introduce LLM and Model Merging. These steps enables cross-dataset attacks using LLMs and combine multiple attackers' knowledge to create a more efficient, generalized model. Specifically, in Attacker Step 1 : Introducing LLM, attacker possesses unique data knowledge, including node features, and can access publicly available pre-trained LLMs. As part of the attack process, attackers act as users by sending their node features to the target model and obtaining posterior probabilities for these nodes. Leveraging the obtained posterior probabilities, the original node features, and a pre-trained LLM, each attacker fine-tunes their LLM to develop an individual attack model capable of performing cross-dataset attacks. In Attacker Step 2: Model Merging, to further enhance the attack's effectiveness, attackers share their fine-tuned LLM parameters and apply a novel model merging technique to integrate their knowledge. This results in a unified and efficient attack model with superior generalization capabilities.

In summary, we introduce LLMs to enable cross-dataset attacks and propose a novel model merging method to integrate the knowledge of multiple attackers. The following sections provide a detailed explanation of the use of LLMs for link stealing attacks and the proposed model merging technique.

\subsection{Introducing LLM for Cross-dataset Link Stealing Attacks} \label{sec: LLM-base}
We introduce LLMs to perform cross-dataset link stealing attacks. LLMs leverage the attention mechanism inherent in their transformer architecture, which is particularly effective at handling variable-length data \cite{DBLP:conf/acl/LiYBZLSLSYWLXBF24}. This flexibility allows LLMs to process diverse features across different datasets, making them highly suitable for cross-dataset link stealing tasks.

To ensure the success of model merging, we standardize the methods and model architectures used by participating attackers for training their LLMs. The process for each standardized attacker to train the LLM mainly involves prompt design and fine-tuning of the LLM.

Prompts are a common mechanism in LLMs and have become the mainstream paradigm for adapting LLMs to specific tasks. With the help of prompts, LLMs can handle various complex tasks. By providing specific prompts to the LLM as input, we can guide the model towards producing the desired output and obtaining the expected results. Similarly, to successfully execute the link stealing attack task using an LLM, designing a customized prompt for this task is crucial.

In many natural language processing tasks, a well-designed prompt specific to the task allows an LLM to produce the expected results without the need for fine-tuning. However, in certain tasks, such as link stealing attacks, the original LLM may not generate the desired results. For example, when attempting a link stealing attack, the LLM's response might be as follows:
\begin{center}
\emph{Based on the information provided, it is not possible to determine if the two papers have a link.}
\end{center}

To effectively integrate link stealing attacks with an LLM, it is necessary to adjust the model. Task-specific prompt fine-tuning is an effective technique for tailoring an LLM to specific tasks \cite{DBLP:journals/csur/LiuYFJHN23}. We employ this technique to optimize the LLM for successful execution of the link stealing attack. 

Next, in this paper, we will detail our approach to Prompt Design and Fine-Tuning of LLM to obtain multiple attack models.

\subsubsection{Prompt Design}
In prompt design, attackers first access the target model using the node features to obtain the posterior probabilities of the corresponding nodes. To facilitate the inference of links between nodes, we group all nodes for which we need to infer the existence of a link into node pairs. The features of each node pair include the node features $\mathrm{X}$ and the posterior probabilities $\mathrm{R}$ of the two nodes, as shown below:

\begin{equation}
    \begin{gathered}
        \mathcal{A}: \underbrace{\{(x_v, r_v), (x_u, r_u)\}}_\mathrm{Node\ Pair},\mapsto \text ( \emph{Link}, \emph{Unlink} )
    \end{gathered}
\end{equation}
where $x_v$ and $x_u$ represent the node features of nodes $v$ and $u$, including textual descriptions. Meanwhile, $p_v$ and $p_u$ represent the posterior probabilities generated by the target model for nodes $v$ and $u$, respectively. Attackers can directly input these node pairs into the attack model to determine whether a link exists between the two nodes in the pair.

Using the obtained node pairs, we design prompts for the LLM. These prompts consist of three main components: i) Node pair information: This includes the node features with text descriptions available to the attacker, as well as the posterior probabilities obtained from the target model. ii) Human question: A natural language query aimed at conducting link stealing attacks based on the node pair information. iii) LLM response: The response generated by the LLM in response to the human question, indicating whether there is a link between the nodes in the pair.

\begin{figure}[htp]
\centering
\includegraphics[scale=0.62]{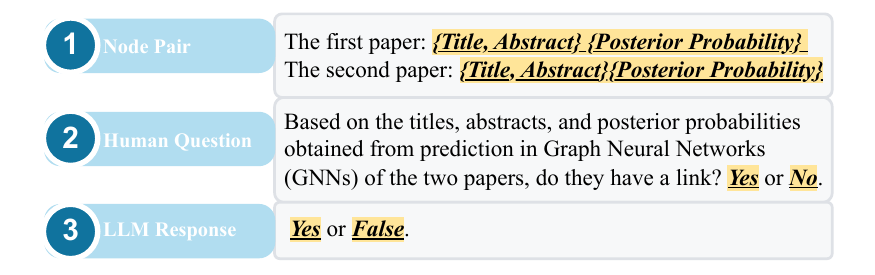}
\caption{Prompts designed for LLM-based link stealing attacks.}
\label{prompt}
\end{figure}

We visualize a prompt example for link stealing attacks, as shown in Fig. \ref{prompt}. This prompt is based on a citation dataset. The node pair information includes the titles and abstracts of two papers, along with the posterior probabilities obtained from the target model. The human question asks the LLM to determine whether a link exists between the two papers based on the information provided. The LLM's response indicates the model's answer: \emph{Yes} signifies that a link exists between the two papers, while \emph{No} indicates that no link.

\subsubsection{Fine-Tuning of LLM to Obtain Multiple Attack Models}
The original LLM cannot effectively complete the link stealing attack task. After designing specific prompts for link stealing attacks, each attacker uses the designed prompt to package the knowledge of individual attackers. Then, each attacker fine-tunes the LLM using these prompts to obtain their individual model capable of performing link stealing attacks. Specifically, we break down the process of each attacker fine-tuning the LLM into the following four steps:

\begin{itemize}

    \item \textbf{Training Data:} The training dataset refers to each attacker data used to fine-tune the LLM but not share. In this paper, the data consists of each attacker well-designed prompts, which are composed of node pair information. This section also covers the tokenization of the training dataset. Tokenization is the process of breaking down text into smaller units known as tokens, which enables the LLM to process and understand the text more effectively \cite{DBLP:conf/acl/SennrichHB16a}.

    \item \textbf{Architecture Supporting Cross-Dataset Attacks:} LLMs typically employ deep neural network architectures, such as the Transformer \cite{DBLP:journals/eswa/IslamEEBDRP24} architecture. Transformers are designed to process sequences of arbitrary lengths, a flexibility enabled by their reliance on attention mechanisms and positional encodings rather than a fixed network structure tied to input size. This capability makes LLM well-suited for cross-dataset link stealing attacks involving datasets with varying data lengths. Common LLMs include Vicuna-7B, Vicuna-13B \cite{DBLP:conf/nips/ZhengC00WZL0LXZ23}, and LongChat \cite{longchat2023}.

    \item \textbf{Generation:} Generation refers to the LLM producing text in response to an input prompt during training or inference. The generated text serves as the answer to the prompt. In this paper, the generation specifically involves determining whether there is a connection between the node pairs in the prompt, resulting in a \emph{Yes} or \emph{No} answer.

    \item \textbf{Fine-tune:} Fine-tuning involves further training a pre-trained model using a small, task-specific dataset to adapt it to a particular application scenario. In link stealing attacks, each attacker compares the predicted output $\hat{\mathrm{Y}}$ generated by the LLM with the true label $\mathrm{Y}$ to compute the loss and fine-tune the model. Each attacker uses cross-entropy $\mathcal{L}_{CE}$ as the loss function, as shown below:

    \begin{equation}
        \begin{gathered}
            \hat{y} = \operatorname{LLM}(\operatorname{Prompt}\{(x_v, p_v), (x_u, p_u)\})\\
            \mathcal{L}_{C E}\left(y, \hat{y}\right) =-\left[y \operatorname { l o g } (\hat{y}+\left(1-y\right) \log \left(1-\hat{y}\right)\right]
        \end{gathered}
    \end{equation}

\end{itemize}

\subsection{Model Merging to Combine Multiple LLM-Based Link Stealing Attack Models}  \label{sec:Federated Learning}
In this paper, to enable each attacker to collaboratively train a more comprehensive link stealing attack model without exchanging data, we propose a novel model merging method. This approach allows us to harness the strengths of each LLM-based link stealing attack model—trained independently on each attacker’s dataset—to create a more effective, unified model. Notably, our method achieves this enhanced performance without requiring data sharing or retraining. The merging process consists of three main steps: Drop, Elect, and Merge. Next, we detail each step of our approach. In this section, we denote $\theta$ as the pre-trained LLM parameters shared among all attackers, and $\{{\theta}^1, {\theta}^2, \dots, {\theta}^n\}$ as the parameters of the LLM attack models obtained by each attacker through fine-tuning based on their respective knowledge. We define the $\delta^t$ parameter as the difference between the parameters of each attack model and the base model, expressed as $\delta^t = {\theta}^t - {\theta}$.

\subsubsection{Drop}

As is well known, many redundant parameters have little to no impact on the performance of a model. LLMs also face the problem of parameter redundancy. In other words, directly pruning these redundant parameters results in minimal decline in model performance. Therefore, we first prune the LLM parameters before merging.

Inspired by \cite{DBLP:journals/tkde/LiMCZLMY23, DBLP:journals/corr/abs-2406-11617}, we employ a pruning technique called Drop to reduce the redundant parameters of the LLM and enhance its generalization ability.

\begin{figure}[htp]
\centering
\includegraphics[scale=0.8]{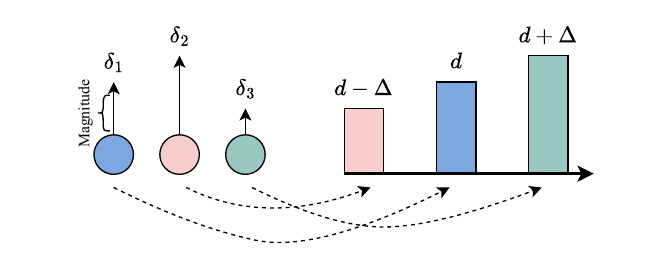}
\caption{Schematic diagram of mapping weights to be inversely proportional to dropout probability.}
\label{drop}
\end{figure}

Drop differs from common random pruning methods, such as Dropout \cite{DBLP:journals/tkde/LiMCZLMY23}. We perform pruning by removing parameters with smaller delta magnitudes, as shown in Fig. \ref{drop}. First, we sort the delta parameters $\{\delta_1, \delta_2, \ldots \delta_n\}$ of the LLM to be merged according to their magnitudes. Based on the sorting results, we assign different drop probabilities, with those having smaller magnitudes receiving higher drop probabilities. The formulation is as follows:

\begin{equation}
    \begin{gathered}
    \left\{r_1, r_2, \ldots, r_n\right\} =\operatorname{rank}\left(\left\{\delta_1, \delta_2, \ldots \delta_n\right\}\right) \\
    \Delta_i =\frac{\epsilon}{n} * r_i \\
    d_i =d_{\min }+\Delta_i
    \end{gathered}
\end{equation}
where $d_{\min}$ is assigned to the maximum magnitude delta parameter as the minimum probability of dropping, defined as $d_{\min} = d - \frac{\epsilon}{2}$. Here, $d$ is the average drop probability, and $\epsilon$ represents the maximum change in drop probability based on magnitude.

By sorting delta magnitudes, we obtain different drop probabilities $d_i$. Next, we use $d_i$ to perform the delta sampling step:
\begin{equation}
    \begin{gathered}
    m_{i} \sim\mathrm{Bernoulli}(d_i) \\
    \tilde{\delta}_{i} =(1-m_i)\odot\delta_i \\
    \hat{\delta}_{i} =\frac{\tilde{\delta}_i}{1-d_i}
    \end{gathered}
\end{equation}
where $ m_i = 1 $ indicates that $ \delta_i $ is dropped, and $m_i = 0$ indicates that it remains. Similar to Dropout \cite{DBLP:journals/tkde/LiMCZLMY23}, to maintain consistency in the final predictions after dropping deltas, we rescale the remaining deltas. Rescaling according to the inverse proportion of the drop $1 - d_i$ can restore the predicted value to its proximity before the drop. We will provide a detailed proof in the subsequent section.

\subsubsection{Elect}
To further reduce the interference from abnormal delta parameters $\delta $ during merging, we perform selective filtering of the delta parameters. We employ a masking technique, where a valid position is represented by $1$ and an invalid or filled position is represented by $0$. By using masks, we can mitigate the influence of anomalies during merging and enhance the effectiveness of the model integration. Specifically, we first calculate the sign of the sum of delta parameters of all models at position $k$ to determine the dominant direction: $ S = \mathrm{sgn} \left( \sum_{t=1}^{T} \hat{\delta}_{k}^{t} \right) $. We then apply the mask to the parameters at position $ k $. The mask value is set to $1$ when the sign of each model's increment aligns with the sign of the overall sum; otherwise, it is set to $0$. The formulation is as follows:

\begin{equation}
    \begin{gathered}
    mask = \begin{cases}1 & \text { if } \mathrm{sgn}(\hat{\delta}_k^t)=S;\\
    0 & \text { else }
    \end{cases}
    \end{gathered}
\end{equation}
where $ T $ is the number of LLM models being merged. After obtaining the mask, we apply it to the delta parameters, resulting in $ {\delta^{\prime}}_k^t = {mask} \odot (\hat{\delta}_k^t) $.

\subsubsection{Merge}
When fusing multiple models, previous methods often rely on calculating the average value of several models for merging \cite{DBLP:journals/corr/abs-2406-11617, DBLP:conf/icml/Yu0Y0L24}. Although this approach can achieve merging, it does not take into account the performance of each individual model, which can lead to inferior models negatively impacting the performance of the merged model.

To address this issue, we propose a new merging strategy that considers each model's accuracy by assigning different weights, denoted as $\lambda$, during the merging process. Higher weights are given to attack models trained on datasets that are more distinct and harder for models trained on other datasets to predict effectively. Specifically, we calculate the accuracy of each LLM-based link stealing attack model trained on a single dataset and evaluated across multiple datasets. By comparing the maximum and minimum accuracies on each dataset, we observe that the maximum accuracy typically occurs when the training and testing datasets are the same, whereas the minimum accuracy is often observed when they differ.

Next, we analyze the difference between the maximum and minimum accuracies. A larger difference indicates that a dataset is more unique and shares fewer similarities with other datasets, making it harder for models trained on other datasets to transfer effectively to it. During model merging, we can increase the merging weights for attack models trained on such distinctive datasets, enhancing the overall generalization ability of the merged model. Finally, we apply the Softmax function to the differences between maximum and minimum accuracies to establish the final weights. This method assigns higher weights to attack models trained on datasets that are challenging to predict and lower weights to those trained on datasets that are easier to predict, thus producing a merged attack model with stronger generalization capabilities. This process can be formulated as:

\begin{equation}
    \begin{gathered}
    {\lambda}^t=\frac{\exp(\text{Max(Acc}^t)-\text{Min(Acc}^t))}{\sum\exp(\text{Max(Acc})-\text{Min(Acc}))}
    \end{gathered}
\end{equation}

\begin{figure}[htp]
\centering
\includegraphics[scale=0.5]{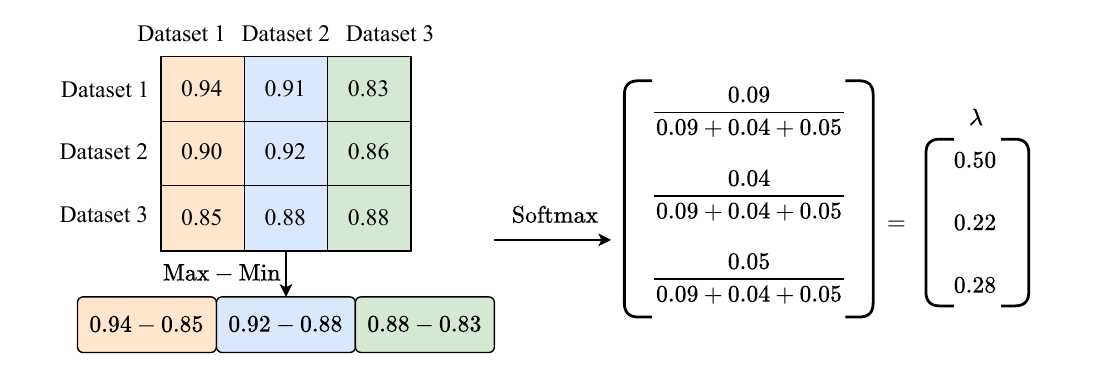}
\caption{Illustrative example of the weight calculation process for $\lambda$.}
\label{softmax}
\end{figure}

We simulate an example to illustrate the calculation process of $\lambda$, as shown in Fig. \ref{softmax}. Each attacker initially trains three attack models on three separate datasets. These models are then evaluated on the same three datasets to obtain their link stealing accuracy values. Next, we calculate the difference in attack accuracy for each dataset by subtracting the minimum accuracy from the maximum accuracy, yielding difference values of $0.09$, $0.04$, and $0.05$. By applying the Softmax function to these values, we derive the $\lambda$ weight values for each attack model during merging: $0.5$, $0.22$, and $0.28$, respectively.

Using the obtained weight $\lambda$, we perform model fusion. Let ${\delta^{\prime}}^{merge}$ denote the merged delta parameter, as shown below:
\begin{equation}
    \begin{gathered}
    {\delta^{\prime}}_k^{merge}={\sum}_{t=1}^{T}{\lambda}^t*{\delta^{\prime}}_k^t
    \end{gathered}
\end{equation}

Finally, we obtain the merged model parameters as
${\theta}_k^{merge}={\theta} + {\delta^{\prime}}_k^{merge}$. By assigning different weights $\lambda$ to models trained on different datasets during the merging process, we create a more reasonable and generalization-capable merged model. This weighted approach ensures that models with better performance contribute more significantly to the final merged model, improving its overall effectiveness and adaptability.

\begin{algorithm}[t]
\caption{The Process of Multiple Attackers Collaborating to Perform Link Stealing Attacks.} \label{alg_extra} 

{\bf Input:} 
    Graph Node Features $X$, Original LLM Parameters $\theta$, Maximum Change $\epsilon$, and Drop Probability $d$\\
{\bf Output:} 
    $Link\ or\ Unlink$
\begin{algorithmic}[1]
\State \textbf{Step 0:} Obtain node posterior probabilities
\State $\ \ \ \ \ \ \mathrm{P} = \mathcal{T}(\mathrm{X})$
\State 
\For {$i = 0$ to $T$} 

    \State \textbf{Step 1:} Designing prompt 
    \State $\ \ \ \ \ \ \operatorname{Prompt}=\{(x_v, p_v), (x_u, p_u)\}$
    \State 

    \State \textbf{Step 2:} Fine-tuning the LLM model using prompt
    \State $\ \ \ \ \ \ {\theta}^t = \operatorname{LLM}({\theta},\operatorname{Prompt})$
    \State 
    
    \State \textbf{Step 3:} Compute delta parameters
    \State $\ \ \ \ \ \ \delta^t = {\theta}^t - {\theta}$
    \State 
    
    \State \textbf{Step 4:} Apply Drop and Scaling
    \State $\ \ \ \ \ \ \hat{\delta}^{t} = \mathrm{Drop\_Scaling}(\delta^t, \epsilon, d)$
    \State 
\EndFor

\State \textbf{Step 5:} Apply mask and election
\State $\ \ \ \ \ \ {\delta^{\prime}} = \mathrm{mask} \odot \hat{\delta}$
\State 

\State \textbf{Step 6:} Merge deltas to obtain merged parameters
\State $\ \ \ \ \ \ {\theta}^{\mathrm{merge}} = {\theta} + \sum_{t=1}^{T}{\lambda}^t \cdot \delta^{\prime t}$ 
\State 

\State \textbf{Step 7:} Link stealing attack 

\State \ \ \ \ \ \ $Link\ or\ Unlink = \operatorname{LLM}
({\theta}^{\mathrm{merge}},\operatorname{Prompt})$\\

\Return $Link\ or\ Unlink$

\end{algorithmic}
\end{algorithm}

\subsection{Link Stealing}
After obtaining the merged model through model merging, we use it to carry out link stealing attacks on the target model. Compared to an attack model trained by a single attacker with limited knowledge, the merged model has stronger generalization capabilities, allowing it to conduct attacks across multiple models more effectively.

The attack process proceeds as follows: First, the attacker sends the features of the two nodes targeted for link stealing to the target model, obtaining the posterior probabilities for these nodes. Next, using a specially designed prompt that combines the node features and posterior probabilities, the attacker generates input features for the link stealing attack model. Finally, the merged model leverages this prompt to infer whether a link exists between the two nodes in question, effectively conducting the link stealing attack.

The algorithm \ref{alg_extra} presents our proposed link stealing attack method and outlines the process for launching attacks. In Step $0$, the attacker accesses the target model to obtain the posterior probabilities of the nodes. Steps $1$–$6$ describe the construction process of our proposed attack method. Specifically, in Step $1$, each attacker first designs the link stealing attack prompts based on their possessed knowledge. Then, in Step $2$, each attacker fine-tunes the original LLM using these prompts to develop their respective attack models. In Step $3$, the attackers calculate the magnitude of the changes in the model parameters before and after fine-tuning. In Steps $4$ and $5$, based on these magnitudes, attackers perform drop, scaling, and elect operations on their model parameters. In Step $6$, the attacker merges the processed changes in the model parameters of each model to obtain the final parameters of the merged model. Step $7$ is the execution phase of the link stealing attack, where the attacker carries out the attack using the merged model and a prompt containing information on the two nodes for which the attacker seeks to steal the link.

\section{Theoretical Analysis}
In this section, we provide a theoretical analysis to explain why the introduction of LLMs facilitates cross-dataset attacks. Additionally, we examine the impact of the Drop and Scaling techniques in model merging on the overall effectiveness of the attack.

\subsubsection{LLMs Enable Cross-Dataset Link Stealing Attacks}
An LLM is trained on extensive and diverse datasets, encompassing textual and structural patterns across various domains. Consequently, its learned representation function, $ \text{LLM(·)} $, exhibits a robust ability to generalize between datasets with different distributions. Formally:

\begin{equation}
    \begin{gathered}
        \text{LLM}(\mathrm{X}) = \text{F}(\text{Trans}(\mathrm{X}))
    \end{gathered}
\end{equation}
where $\text{Trans(·)}$ refers to the transformer architecture \cite{DBLP:journals/eswa/IslamEEBDRP24} within the LLM. The transformer architecture enables the model to dynamically focus on different parts of the input sequence, regardless of its length. It tokenizes the input data; for example, in natural language processing, a sentence is decomposed into individual words or sub-word units (tokens). Each input token attends to all other tokens, capturing global dependencies without being constrained by sequence length. As a result, the LLM can effectively handle variable data lengths. Additionally, $\text{F(·)}$ represents other neural networks incorporated into the model.

Traditional link stealing attacks lack pre-training on diverse data and depend entirely on supervised learning within a single dataset. The representation function of an MLP, $ \text{MLP(·)} $, is typically expressed as:

\begin{equation}
    \begin{gathered}
        \text{MLP}(\mathrm{X}) =  \sigma(\mathrm{WX} + \mathrm{B})
    \end{gathered}
\end{equation}
where $\mathrm{W}$ is a learned weight matrix of size $k \times n$ (for $n$-dimensional inputs). If $\mathrm{X} \in \mathbb{R}^m$ with $m \neq n$, the matrix multiplication $\mathrm{WX}$ becomes invalid. Thus, an MLP cannot handle the issue of variable data lengths. Additionally, $\mathrm{B}$ represents the bias term, and $\sigma$ is the activation function.

It should be noted that LLMs have achieved state-of-the-art performance in various fields \cite{DBLP:conf/acl/LiYBZLSLSYWLXBF24}. By effectively leveraging the textual features of the nodes, LLMs can significantly enhance the performance of link stealing attacks. In previous link stealing attacks, methodological limitations prevented researchers from using the textual and structural features of the nodes. Therefore, introducing LLMs for link stealing attacks offers the potential to surpass traditional methods in terms of performance.

Based on the analysis and assumptions above, we conclude that introducing LLMs enables cross-dataset attacks and markedly improves attack performance.

\subsubsection{Drop and Scaling Maintain Attack Effectiveness}
Here, we analyze the efficiency of the Drop and Scaling techniques in merging multiple models. First, let us define the model parameters as $\mathrm{W} $ and the corresponding deltas as $ \delta $. Let $ \mathrm{X} $ be the input embedding vector and $ \mathrm{H} $ represent the output embedding. The model's predictions can be formulated as:
\begin{equation}
    \begin{aligned}
    \mathbb{E}[h]& =\mathbb{E}\left[\sum_{i=1}^n\left(w_i+\delta_i\right)x_i\right] \\
    &=\sum_{i=1}^nx_i\mathbb{E}[w_i]+\sum_{i=1}^nx_i\mathbb{E}[\delta_i] \\
    &=\sum_{i=1}^nx_iw_i+\sum_{i=1}^nx_i\delta_i=h+\Delta h
    \end{aligned}
\end{equation}

Let $ d $ denote the probability of drop in the drop and scaling process and let $ \gamma$ represent the scaling factor. After applying Drop and Scaling to the model's parameters, the predictions can be formulated as:

\begin{equation}
    \begin{aligned}
    \mathbb{E}[\hat{h}]& =\mathbb{E}\left[\sum_{i=1}^n\left(w_i+\hat{\delta}_i\right)x_i\right] \\
    &=\sum_{i=1}^nx_i\mathbb{E}[w_i]+\sum_{i=1}^nx_i\mathbb{E}[\hat{\delta}_i] \\
    &=\sum_{i=1}^nx_iw_i+\sum_{i=1}^nx_i(1-d_i)*\gamma*\delta i \\
    &=h+\sum_{i=1}^nx_i*(1-d_i)*\gamma*\delta_i
    \end{aligned}
\end{equation}

By setting $ \gamma = \frac{1}{1 - d_i} $, we ensure $\mathbb{E}[\hat{h}]=\mathbb{E}[h]$, thereby concluding that the expected value of the rescaled output matches the original prediction before applying the Drop and Scaling procedure. This ensures the consistency of the model's predictions. Specifically, we can express this as follows:
\begin{equation}
    \begin{aligned}
    \mathbb{E}[\hat{h}]&=h+\sum_{i=1}^nx_i*\delta_i\\
    \mathbb{E}[\hat{h}]&=h+\Delta h\\
    \mathbb{E}[\hat{h}]&=\mathbb{E}[h]\end{aligned}
\end{equation}

Based on this theoretical analysis, we can infer that applying Drop and Scaling to the LLM in a link stealing attack does not impair the attack's performance.

\section{Experimental Evaluations}
\subsection{Experimental Setup}

\subsubsection{Datasets}
To verify the effectiveness of the proposed method, we conducted experiments on four datasets: Cora, Citeseer, Pubmed \cite{DBLP:conf/iclr/KipfW17}, and Ogbn-arxiv \cite{zheng2021grb}. Cora, Citeseer, and Pubmed are widely used in Graph Neural Network (GNN) research, allowing effective comparisons with existing methods. We included Ogbn-arxiv, a larger dataset, to align with the current trend of increasing dataset sizes in research. Table \ref{dataset} provides an overview of these dataset statistics.

\begin{table}[htbp]
  \centering
  \caption{Dataset Statistics. \textbf{Attack Links} indicates the number of links known by the attacker.}
    \begin{tabular}{c|ccccm{1cm}<{\centering}}
    \toprule
          & \textbf{Nodes} & \textbf{Feats} & \textbf{Links} & \textbf{Classes} & \textbf{Attack Links}\\
    \midrule
    \textbf{Cora} & $2,708$ & $1,433$ & $10,556$ & $7$ &$2,000$\\
    \textbf{Citeseer} & $3,327$ & $3,703$ & $9,228$ & $6$ &$2,000$\\
    \textbf{Pubmed} & $19,717$ & $500$   & $88,651$ & $3$ &$5,000$\\
    \textbf{Ogbn-arxiv} & $169,343$ & $128$   & $1,166,243$ & $40$ &$30,000$\\
    \bottomrule
    \end{tabular}%
  \label{dataset}%
\end{table}%

\subsubsection{Dataset Configuration}
In link stealing attacks, we assume that attackers have knowledge of a varying number of links based on the dataset size. As shown in Table \ref{dataset}, the \emph{Attack Links} column specifies the number of links known to the attacker. Specifically, attackers are aware of $2,000$ links in the Cora and Citeseer datasets, $5,000$ in Pubmed, and $30,000$ in Ogbn-arxiv. To ensure fairness during model training, an equal number of unlinked edges are randomly selected for inclusion in the training process.

\subsubsection{Models}
To explore the effectiveness of our proposed method, we used Vicuna-7B \cite{DBLP:conf/nips/ZhengC00WZL0LXZ23} to conduct link stealing attacks against the Graph Convolutional Network (GCN) \cite{DBLP:conf/iclr/KipfW17}, a widely adopted Graph Neural Network (GNN) model. Vicuna-7B is also a prominent large language model (LLM) used for various tasks.

Additionally, to assess the versatility of our method, we extended our experiments to include multiple GNN models and LLM architectures. The GNN models tested included GCN, Graph Attention Networks (GAT) \cite{DBLP:conf/iclr/VelickovicCCRLB18}, and GraphSAGE (SAGE) \cite{DBLP:conf/nips/HamiltonYL17}, which are popular in GNN research. Similarly, we evaluated the method using several LLM architectures, including Vicuna-7B, Vicuna-13B \cite{DBLP:conf/nips/ZhengC00WZL0LXZ23}, and LongChat \cite{longchat2023}. This comprehensive testing demonstrates the robustness and generalizability of our approach across a range of models.

\subsubsection{Evaluation Metric}
We use Accuracy and F1 score to evaluate the quality of the model. Accuracy refers to the proportion of samples correctly predicted by the model out of the total number of samples. It reflects the overall prediction accuracy, i.e., the ratio of correct predictions to all predictions. 
The F1 score is the harmonic mean of precision and recall, providing a balanced measure of the model's performance, especially useful for imbalanced datasets. 

\subsection{Evaluation of the Attack Model Trained Under a Single Dataset}
We first explore the effectiveness of our proposed LLM-based Link Stealing Attacks in scenarios where the attacker only has access to links from a single dataset. In this setting, the attacker possesses partial link information from a single dataset and does not know the links in other datasets.

\subsubsection{Effectiveness of LLM-based Link Stealing Attacks}
Table \ref{effect} presents the experimental results of our proposed LLM-based link stealing attack method across four datasets: Cora, Citeseer, Pubmed, and Ogbn-arxiv. In the table, the \emph{Feature} method indicates that the attacker uses node features alone to carry out the attack. \emph{PP} represents attacks carried out using posterior probabilities of the target model. \emph{PP+Feature} combines both node features and posterior probabilities for the attack. \emph{LLM (Our)} refers to the results obtained by our proposed LLM-based link stealing attack method, trained on a single dataset.

As shown in the table, the LLM-based link stealing attack method proposed in this paper outperforms previous methods in both accuracy and F1 score. Specifically, our method improves accuracy by at least $3\%$ and F1 score by at least $2\%$ across the datasets. On the Pubmed dataset, the improvement is most significant, with a $7\%$ increase in both accuracy and F1. Additionally, on the Ogbn-arxiv dataset, our method achieves remarkable results, with an accuracy of $97.48\%$ and an F1 score of $97.49\%$.

\begin{figure}[tbp]
  \centering
  \subfigure[Accuracy]{
  \begin{minipage}[b]{0.23\textwidth}
    \includegraphics[width=\textwidth]{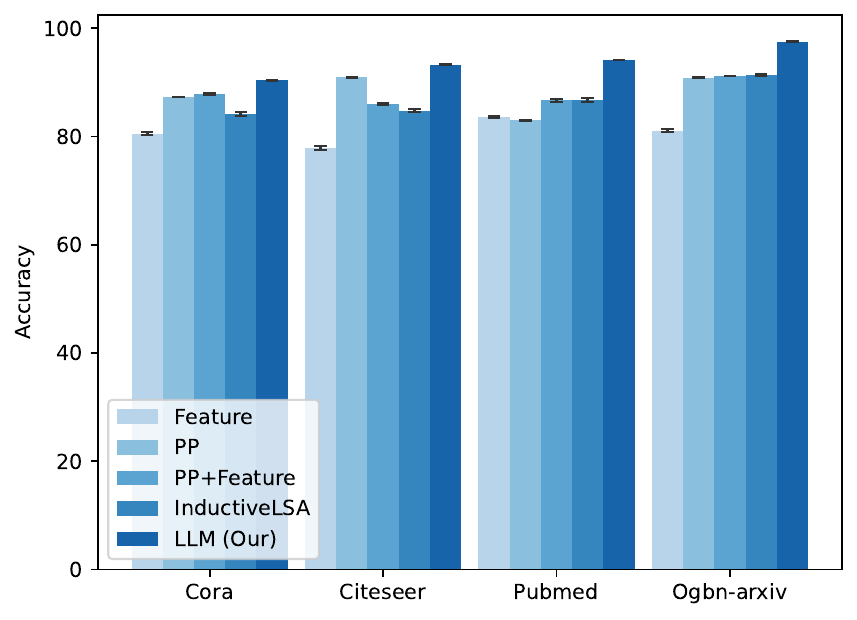}
  \end{minipage}
  }
  \subfigure[F1]{
  \begin{minipage}[b]{0.23\textwidth}
    \includegraphics[width=\textwidth]{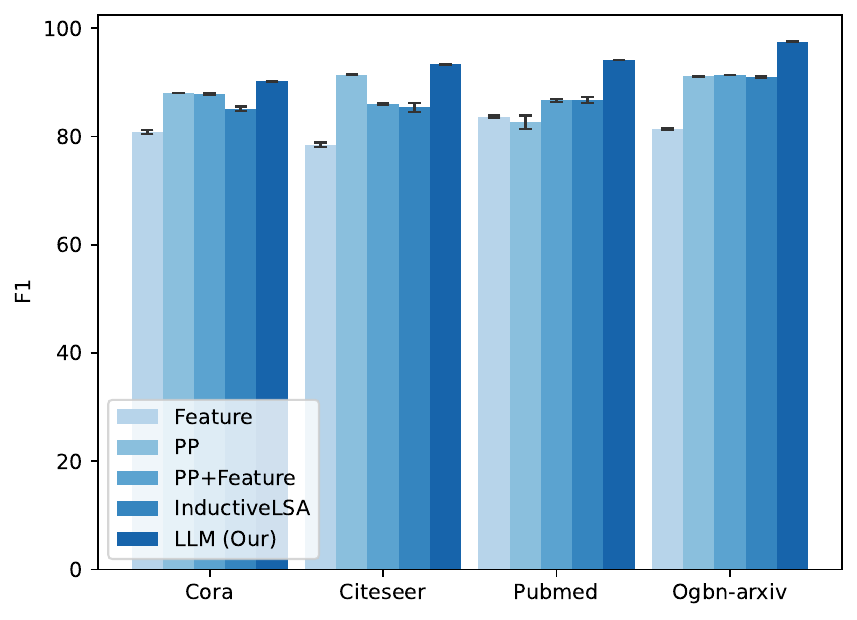}
  \end{minipage}
  }

  \caption{Comparison of different link stealing attack methods trained under a single dataset.}
  \label{com_all}
\end{figure}

To provide a more intuitive understanding of the superiority of our method, we visualized the comparison results in Fig. \ref{com_all}. From both Table \ref{effect} and Fig. \ref{com_all}, it is clear that the LLM-based link stealing attack method we proposed significantly outperforms previous approaches. This shows that our method effectively combines the textual features of the nodes with the traditional link stealing attack features, thus improving the success rate of the attacks.

\subsubsection{Evaluation of Cross-Dataset Link Stealing Attacks}
\begin{figure}[tbp]
  \centering
  \subfigure[Accuracy]{
  \begin{minipage}[b]{0.23\textwidth}
    \includegraphics[width=\textwidth]{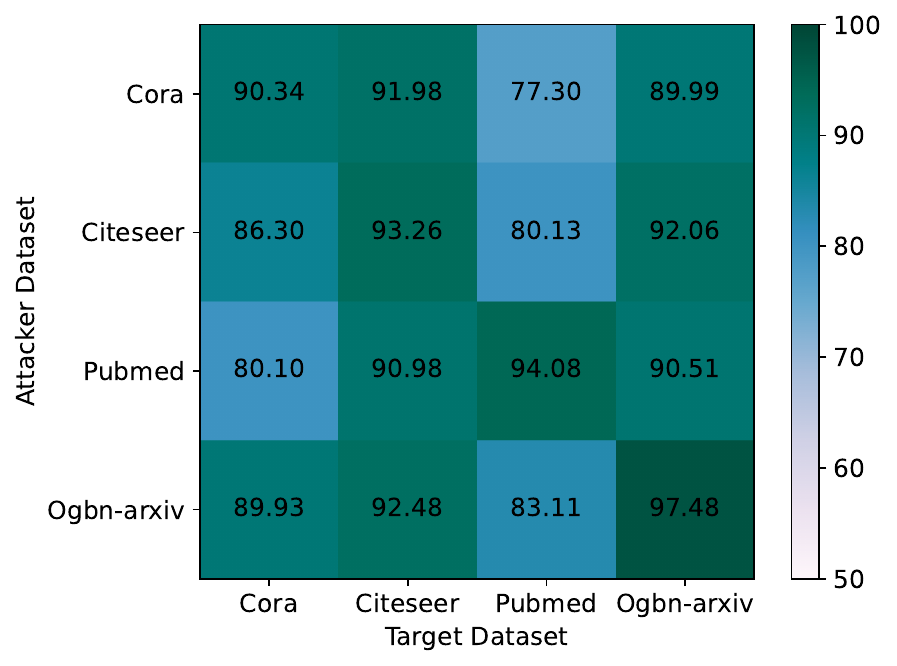}
  \end{minipage}
  }
  \subfigure[F1]{
  \begin{minipage}[b]{0.23\textwidth}
    \includegraphics[width=\textwidth]{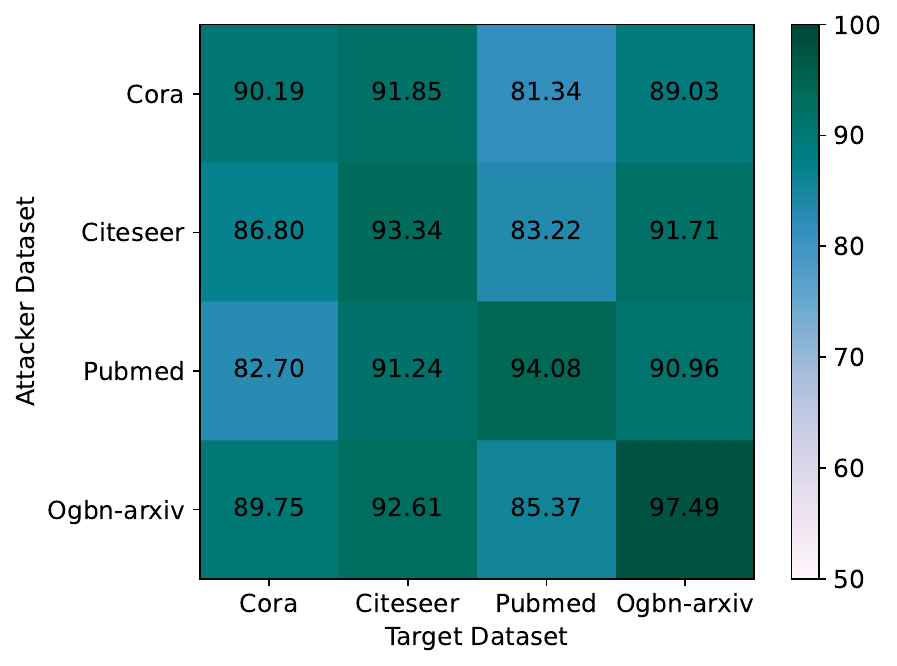}
  \end{minipage}
  }

  \caption{Heat map of LLM performance in LLM-base link stealing attacks across different datasets}
  \label{heat_map}
\end{figure}

Previous link stealing attack methods \cite{DBLP:conf/uss/HeJ0G021,DBLP:journals/popets/WuHBHBGZ24} struggle to perform cross-domain dataset attacks because the attack features (posterior probabilities) differ in dimensions across datasets. Consequently, a single model can only carry out attacks on a single dataset. In contrast, the LLM-based link stealing attack method handles variable-length data features effectively, overcoming this limitation and enabling cross-dataset attacks. To test this, we applied the LLM attack model trained on a single dataset to attack all datasets. The experimental results are shown in Fig. \ref{heat_map}.

As shown in the table, the LLM-based link stealing attack model, trained on a single dataset (Cora, Citeseer, Pubmed, or Ogbn-arxiv), can successfully attack all four datasets simultaneously. Notably, in some cases, attacks using a model trained on a different dataset outperform previous non-LLM methods trained and tested on the same dataset. For instance, the LLM attack model trained on Ogbn-arxiv achieves an accuracy of $89.93\%$ on Cora and $92.48\%$ on Citeseer, surpassing the previous methods' accuracy of $87.83\%$ and $90.91\%$, respectively. This shows that our proposed LLM-based link stealing attack method can be generalized across different datasets and still achieve superior results.

However, due to the inherent differences among various datasets, using an attack model trained on one dataset to attack another with significant differences may not yield optimal results. For example, the attack model trained on Cora achieves $90.34\%$ accuracy on Cora but only $77.3\%$ on Pubmed, which is clearly a suboptimal performance. To address this issue, this paper introduces a model merging approach, enabling the creation of a comprehensive attack model with strong generalization capabilities. This method allows for joint training of the model without requiring multiple attackers to interact with different datasets, ultimately enhancing the effectiveness of the attack across diverse datasets.

\begin{table*}[htbp]
  \centering
  \caption{Results of different link stealing attacks trained under a single dataset. The Feature method indicates that the attacker uses node features alone to conduct the attack. PP represents attacks conducted using posterior probabilities from the target model.}
    \begin{tabular}{c|cccc|cccc}
    \toprule
          & \multicolumn{4}{c|}{\textbf{Accuracy}} & \multicolumn{4}{c}{\textbf{F1}} \\
\cmidrule{2-9}          & \textbf{Cora} & \textbf{Citeseer} & \textbf{Pubmed} & \textbf{Ogbn-arxiv} & \textbf{Cora} & \textbf{Citeseer} & \textbf{Pubmed} & \textbf{Ogbn-arxiv} \\
    \midrule
    \textbf{Feature \cite{DBLP:conf/uss/HeJ0G021}} & 80.51±0.29 & 77.84±0.36 & 83.65±0.18 & 81.08±0.24 & 80.78±0.36 & 78.50±0.39 & 83.66±0.23 & 81.42±0.16 \\
    \textbf{PP \cite{DBLP:conf/uss/HeJ0G021}} & 87.29±0.08 & 90.91±0.09 & 82.97±0.14 & 90.85±0.05 & 88.03±0.05 & 91.41±0.08 & 82.68±1.21 & 91.11±0.04 \\
    \textbf{PP+Feature} & 87.83±0.15 & 85.94±0.19 & 86.65±0.28 & 91.18±0.04 & 87.92±0.17 & 86.04±0.15 & 86.71±0.29 & 91.36±0.03 \\
    \textbf{\textcolor{black}{InductiveLSA \cite{DBLP:journals/popets/WuHBHBGZ24}}} & \textcolor{black}{84.16±0.38} & \textcolor{black}{84.76±0.26} & \textcolor{black}{86.73±0.41} & \textcolor{black}{91.28±0.17} & \textcolor{black}{85.12±0.43} & \textcolor{black}{85.38±0.77} & \textcolor{black}{86.70±0.53} & \textcolor{black}{91.08±0.17} \\
    \textbf{LLM (Our)} & \textbf{90.34±0.08} & \textbf{93.26±0.04} & \textbf{94.08±0.02} & \textbf{97.48±0.08} & \textbf{90.19±0.08} & \textbf{93.34±0.04} & \textbf{94.08±0.02} & \textbf{97.49±0.09} \\
    \bottomrule
    \end{tabular}%
  \label{effect}%
\end{table*}%

\subsection{The Allocation of the Weight Factor $\lambda$ During Model Merge}

In Section \ref{sec:Federated Learning}, we discussed that during model merging, different weights are allocated based on the varying accuracy of each attack model. Table \ref{Allocation} presents the accuracy results of the LLM-based method when trained on a single dataset—Cora, Citeseer, Pubmed, or Ogbn-arxiv—and evaluated across all four datasets. In this table, \emph{Cora-LLM} refers to the attack model trained using Cora as the training set. These accuracy scores guide the weight allocation, allowing the final merged model to maximize performance by drawing on the strengths of the top-performing models for each dataset.

To determine weight allocation during model merging, we first identify the maximum accuracy for attacks on each dataset. Generally, an attack model performs best when its training set matches its test set. Table \ref{Allocation} shows the highest accuracy rates for attacks on Cora, Citeseer, Pubmed, and Ogbn-arxiv, achieved when each dataset was also used as the training set, reaching values of $90.34\%$, $93.26\%$, $94.08\%$, and $97.48\%$, respectively. Next, we find the minimum attack performance on each dataset, allowing us to compute the range (MAX-MIN). This range reveals variations in attack difficulty across datasets. For instance, the Pubmed dataset achieves $94.08\%$ accuracy under the Pubmed-trained LLM attack model, while the same model trained on Cora yields only $77.3\%$ on Pubmed—a difference of $16.78\%$. This highlights Pubmed’s challenge level, as models trained on other datasets struggle to achieve comparable performance. Therefore, in model merging, we can assign a higher weight to the Pubmed-LLM model, as it effectively addresses a more challenging dataset. Conversely, the Citeseer dataset exhibits more consistency across models: accuracy remains around $90.98\%$, regardless of the training dataset, showing only a $2.28\%$ difference from the Citeseer-trained model. This suggests Citeseer’s lower attack complexity, and therefore, the Citeseer-LLM weight can be reduced.

\begin{figure}[htp]
\centering
\includegraphics[scale=1.3]{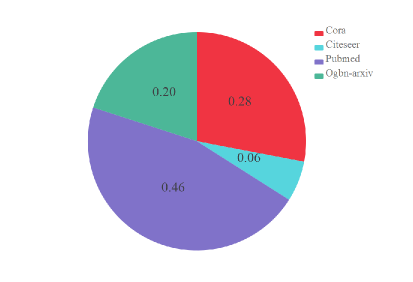}
\caption{Allocation of weight $\lambda$ for each attack model during model merging.}
\label{accuracy_lambda}
\end{figure}

By using MAX-MIN values, we assign higher weights to models trained on more challenging datasets, enhancing overall merge model performance, including generalization for unseen data. Finally, applying a Softmax operation to the MAX-MIN results yields final weights of $0.28$, $0.06$, $0.46$, and $0.2$, as shown in Fig. \ref{accuracy_lambda}, optimizing the robustness of the merged model across datasets.

\begin{table}[htbp]
  \centering
  \caption{Results of calculating allocation weights for the merged model based on accuracy from LLM-based training across four attackers on their respective datasets. \emph{Max} denotes the highest accuracy achieved on the dataset, \emph{Min} represents the lowest accuracy, and \emph{Softmax($\lambda$)} indicates the weights ($\lambda$) computed using the Softmax function.}
    \begin{tabular}{c|cccc}
    \toprule
          & \textbf{Cora} & \textbf{Citeseer} & \textbf{Pubmed} & \textbf{Ogbn-arxiv} \\
    \midrule
    \textbf{Cora-LLM} & 90.34 & 91.98 & 77.3  & 89.99 \\
    \textbf{Citeseer-LLM} & 86.3  & 93.26 & 80.13 & 92.06 \\
    \textbf{Pubmed-LLM} & 80.1  & 90.98 & 94.08 & 90.51 \\
    \textbf{Ogbn-arxiv-LLM} & 89.93 & 92.48 & 83.11 & 97.48 \\
    \midrule
    \textbf{Max} & 90.34 & 93.26 & 94.08 & 97.48 \\
    \textbf{Min} & 80.1  & 90.98 & 77.3  & 89.99 \\
    \textbf{Max-Min} & 10.24  & 2.28   & 16.78 & 7.49 \\
    \midrule
    \textbf{Softmax ($\lambda$)} & 0.28 & 0.06 & 0.46 & 0.2 \\
    \bottomrule
    \end{tabular}%
  \label{Allocation}%
\end{table}%

\subsection{Effectiveness Analysis of the Proposed Merging Method}
Here, we examine the effectiveness of our proposed merging method for one-shot federated learning. We demonstrate its advantages by comparing it with established merging methods, including Ties \cite{DBLP:conf/nips/YadavTCRB23}, Dare \cite{DBLP:conf/icml/Yu0Y0L24}, and Della \cite{DBLP:journals/corr/abs-2406-11617}. The experimental results are presented in Table \ref{effection_merge}.

As shown in Table \ref{effection_merge}, our method achieves superior merging effectiveness in link stealing attacks compared to other approaches. The Mean method, the simplest merging technique, averages the parameters of the models, initially facilitating model merging with an accuracy of $89.83\%$ and an F1 score of $89.87\%$, demonstrating the viability of link stealing attacks for model merging. However, it shows significant performance loss, especially on the Pubmed and Ogbn-arxiv datasets, where accuracy drops exceed $6\%$.

Advanced methods like Ties, Dare, and Della improve merging performance over Mean, particularly benefiting datasets with lower attack accuracy, such as Cora and Citeseer. For instance, $\mathrm{Della\_ties}$ achieves accuracy close to that of individual attack models trained specifically on Cora and Citeseer. However, similar to Mean, these methods experience a $1\%$-$3\%$ accuracy decrease on Pubmed and Ogbn-arxiv. 

\begin{figure}[tbp]
  \centering
  \subfigure[Accuracy]{
  \begin{minipage}[b]{0.23\textwidth}
    \includegraphics[width=\textwidth]{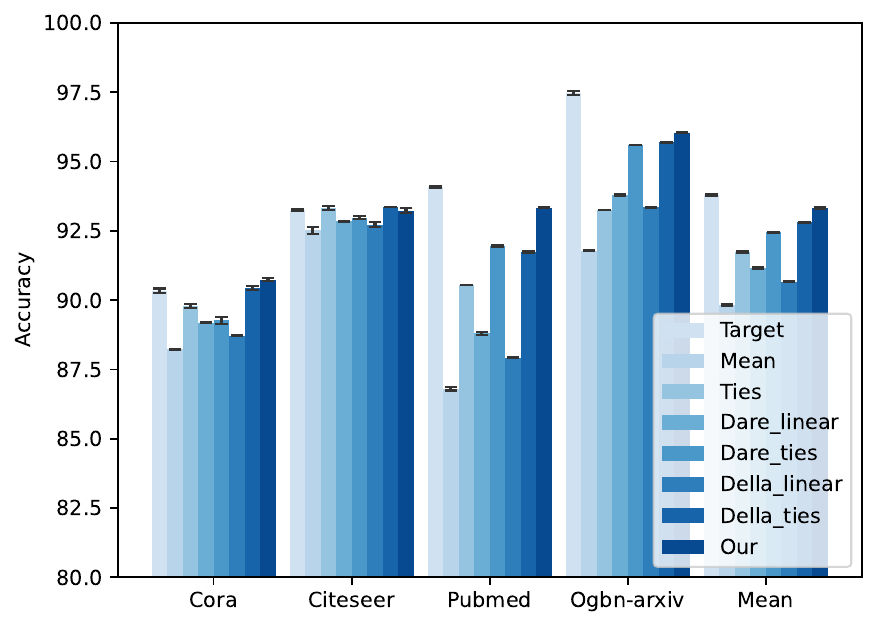}
  \end{minipage}
  }
  \subfigure[F1]{
  \begin{minipage}[b]{0.23\textwidth}
    \includegraphics[width=\textwidth]{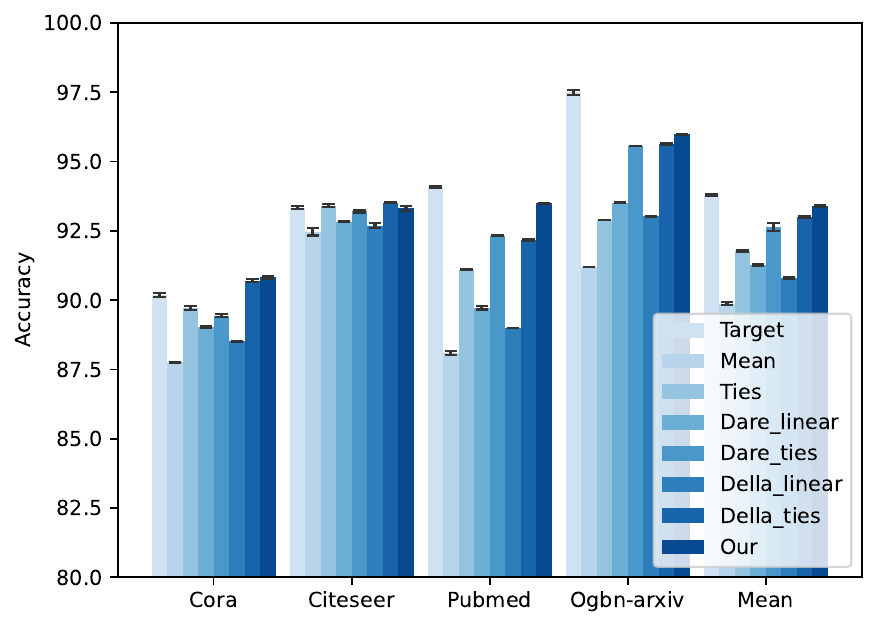}
  \end{minipage}
  }

  \caption{Comparison of model merging method in link stealing attack.}
  \label{com_merge}
\end{figure}

In contrast, our method more effectively preserves prediction performance across all datasets. Even for Pubmed and Ogbn-arxiv, where baseline merged models perform relatively well, our method maintains competitive performance, with only a $0.8\%$ decrease on Pubmed and $1.4\%$ on Ogbn-arxiv. On average, across the four datasets, our method achieves the highest overall accuracy of $93.33\%$ and an F1 score of $93.40\%$, surpassing all existing methods. To visually illustrate the superiority of our method, we provide a comparative column chart of all existing methods in Fig. \ref{com_merge}.

\begin{table*}[htbp]
  \centering
  \caption{Results of different model merging methods for link stealing attacks. \emph{Mean} represents the average attack performance across four different datasets.}
  \scalebox{0.9}{
    \begin{tabular}{c|ccccc|ccccc}
    \toprule
          & \multicolumn{5}{c|}{\textbf{Accuracy}} & \multicolumn{5}{c}{\textbf{F1}} \\
\cmidrule{2-11}          & \textbf{Cora} & \textbf{Citeseer} & \textbf{Pubmed} & \textbf{Ogbn-arxiv} & \textbf{Mean} & \textbf{Cora} & \textbf{Citeseer} & \textbf{Pubmed} & \textbf{Ogbn-arxiv} & \textbf{Mean} \\
    \midrule
    \textbf{Target (Max)} & 90.34±0.08 & 93.26±0.04 & 94.08±0.02 & 97.48±0.08 & 93.79±0.04 & 90.19±0.08 & 93.34±0.04 & 94.08±0.02 & 97.49±0.09 & 93.78±0.04 \\
    \textbf{Mean} & 88.22±0.02 & 92.52±0.12 & 86.81±0.07 & 91.78±0.01 & 89.83±0.04 & 87.75±0.01 & 92.46±0.13 & 88.09±0.06 & 91.20±0.01 & 89.87±0.05 \\
    \textbf{Ties} & 89.80±0.07 & 93.33±0.07 & 90.54±0.00 & 93.25±0.01 & 91.73±0.04 & 89.72±0.08 & 93.41±0.06 & 91.10±0.01 & 92.89±0.01 & 91.78±0.04 \\
    \textbf{Dare\_linear} & 89.19±0.01 & 92.84±0.01 & 88.80±0.06 & 93.80±0.03 & 91.16±0.02 & 89.04±0.04 & 92.83±0.01 & 89.72±0.06 & 93.52±0.03 & 91.28±0.03 \\
    \textbf{Dare\_ties} & 89.27±0.12 & 92.97±0.05 & 91.96±0.03 & 95.58±0.00 & 92.44±0.02 & 89.44±0.60 & 93.21±0.05 & 92.33±0.02 & 95.56±0.00 & 92.63±0.14 \\
    \textbf{Della\_linear} & 88.72±0.02 & 92.72±0.09 & 87.92±0.01 & 93.35±0.01 & 90.67±0.02 & 88.51±0.01 & 92.68±0.09 & 89.00±0.00 & 93.02±0.01 & 90.80±0.02 \\
    \textbf{Della\_ties} & 90.44±0.06 & \textbf{93.35±0.00} & 91.74±0.04 & 95.69±0.03 & 92.81±0.02 & 90.70±0.06 & \textbf{93.51±0.01} & 92.16±0.04 & 95.63±0.03 & 93.00±0.02 \\
    \textbf{Our} & \textbf{90.74±0.06} & 93.22±0.09 & \textbf{93.34±0.02} & \textbf{96.04±0.01} & \textbf{93.33±0.04} & \textbf{90.83±0.05} & 93.32±0.09 & \textbf{93.49±0.02} & \textbf{95.98±0.01} & \textbf{93.40±0.04} \\
    \bottomrule
    \end{tabular}%
    }
  \label{effection_merge}%
\end{table*}%

\subsection{Performance Analysis of the Merging Method Using Different Parameters}
In model merging, two critical hyperparameters are involved. The first is $ p $, which represents the average drop probability, and the second is $ \epsilon $, indicating the maximum allowable variation in drop probability based on parameter magnitude. Here, we explore how these two hyperparameters affect the performance of model merging.

\subsubsection{Effect of Varying Drop Probability $ p $}
To analyze the impact of varying drop probabilities $ p $ on the merged model’s performance, we conducted experiments with $ p = [0.1, 0.3, 0.5, 0.7, 0.9] $, as illustrated in Fig. \ref{drop}. The accuracy and F1 scores shown in the figure represent the average performance of the merged model across four datasets.

As shown in Fig. \ref{drop}, both excessively high or low drop probabilities $ p $ reduce the performance of the merged model. When the drop probability $ p $ is too low, such as at $ p = 0.1 $, model merging fails to achieve optimal performance due to parameter redundancy. LLMs contain numerous finely tuned parameters, leading to significant redundancy \cite{DBLP:conf/icml/Yu0Y0L24}. If these redundant parameters are not pruned before merging, they can skew the parameter calculations and ultimately degrade the performance of the merged model.

\begin{figure}[htp]
\centering
  \subfigure[Drop Probabilities $d$]{
  \begin{minipage}[b]{0.23\textwidth}
    \includegraphics[width=\textwidth]{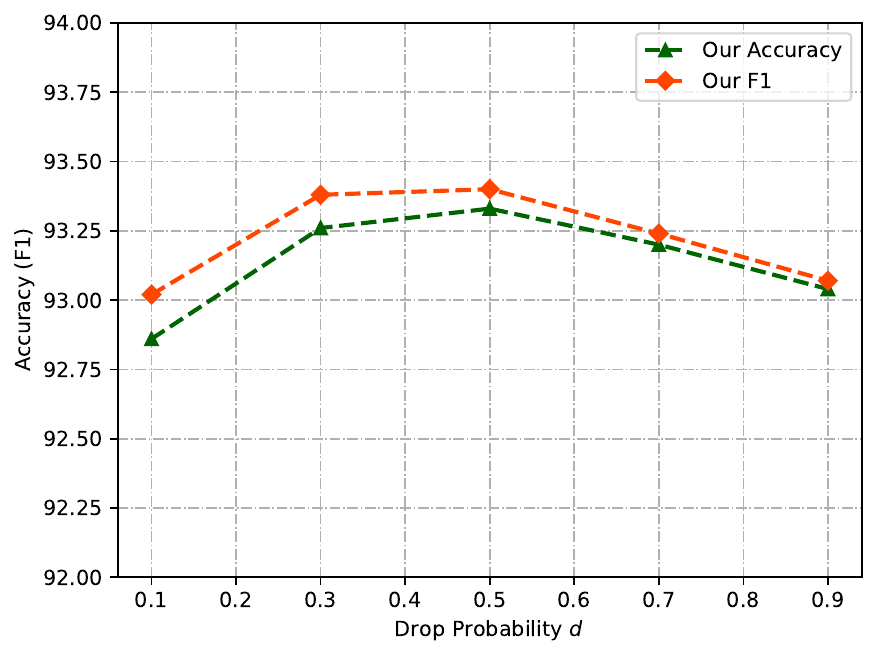}
  \end{minipage}
  }
  \subfigure[Maximum Magnitude $\epsilon$]{
  \begin{minipage}[b]{0.23\textwidth}
    \includegraphics[width=\textwidth]{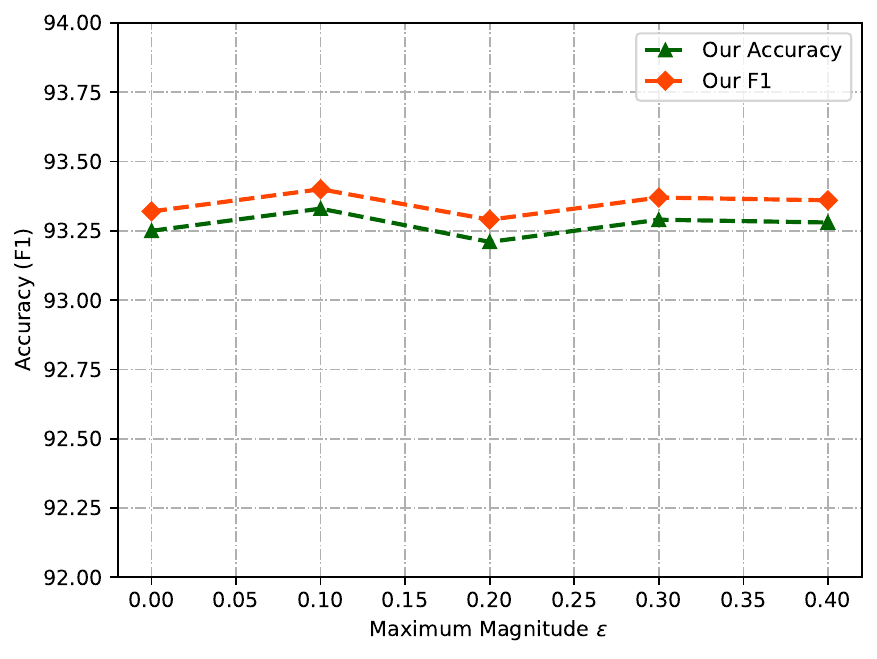}
  \end{minipage}
  }
  \caption{Impact of different drop probabilities $d$ and maximum magnitude $\epsilon$ on model merging results.}
\label{Magnitude}
\end{figure}

Conversely, if the drop probability $ p $ is set too high, such as at $0.7$ or $0.9$, an excessive number of parameters are removed. This aggressive pruning eliminates some essential parameters, resulting in decreased model performance. Based on these observations, we present a comparative figure of this behavior in Fig. \ref{Magnitude} (a). Selecting an appropriate drop probability $ p $ is essential for optimal model merging outcomes. In this paper, unless otherwise specified, we set $ p = 0.5 $, which demonstrates the best performance in figure, for model merging.

\begin{table*}[htbp]
  \centering
  \caption{Results of the proposed attack method using different LLM architectures for link stealing attacks. \textbf{Bold} text highlights the best performance for each dataset, while \underline{underlined} text indicates the second-best performance.}
      \scalebox{0.83}{
    \begin{tabular}{cc|ccccc|ccccc}
    \toprule
          & \multirow{2}[4]{*}{} & \multicolumn{4}{c}{\textbf{Accuracy}} &       & \multicolumn{4}{c}{\textbf{F1}} &  \\
\cmidrule{3-12}          &       & \textbf{Cora} & \textbf{Citeseer} & \textbf{Pubmed} & \textbf{Ogbn-arxiv} & \textbf{Mean} & \textbf{Cora} & \textbf{Citeseer} & \textbf{Pubmed} & \textbf{Ogbn-arxiv} & \textbf{Mean} \\
    \midrule
    \multirow{5}[2]{*}{\textbf{Vicuna-13B}} & \textbf{Cora-LLM} & \textbf{90.62±0.02} & 92.09±0.08 & 87.04±0.01 & 86.91±0.01 & 89.16±0.02 & \textbf{90.58±0.00} & 91.88±0.08 & 88.24±0.00 & 85.05±0.01 & 88.94±0.02 \\
          & \textbf{Citeseer-LLM} & 87.48±0.04 & \textbf{93.31±0.25} & 86.15±0.04 & 89.22±0.02 & 89.04±0.05 & 88.07±0.04 & \textbf{93.36±0.24} & 87.55±0.03 & 88.08±0.03 & 89.26±0.05 \\
          & \textbf{Pubmed-LLM} & 84.72±0.02 & 91.77±0.05 & \underline{92.80±0.08} & 92.23±0.00 & 90.38±0.04 & 86.23±0.01 & 91.87±0.06 & \textbf{93.08±0.08} & 91.91±0.01 & 90.77±0.04 \\
          & \textbf{Ogbn-arxiv-LLM} & \underline{89.76±0.11} & 91.54±0.16 & 91.08±0.01 & \textbf{95.51±0.02} & \underline{91.97±0.07} & 89.98±0.08 & 91.23±0.16 & 91.14±0.02 & \textbf{95.35±0.02} & \underline{91.92±0.06} \\
          & \textbf{Our} & 89.62±0.19 & \underline{93.24±0.08} & \textbf{92.97±0.03} & \underline{94.93±0.01} & \textbf{92.69±0.02} & \underline{90.13±0.19} & \underline{93.25±0.09} & \underline{93.03±0.03} & \underline{94.75±0.01} & \textbf{92.79±0.02} \\
    \midrule
    \multirow{5}[2]{*}{\textbf{Longchat}} & \textbf{Cora-LLM} & \underline{88.13±0.07} & 89.85±0.07 & 87.23±0.06 & 89.66±0.04 & 88.72±0.03 & \underline{87.99±0.09} & 89.57±0.06 & 87.67±0.05 & 88.92±0.04 & 88.54±0.04 \\
          & \textbf{Citeseer-LLM} & 86.07±0.23 & \underline{91.44±0.01} & 86.86±0.15 & 91.16±0.08 & 88.88±0.01 & 86.49±0.22 & \underline{91.50±0.04} & 87.52±0.13 & 90.71±0.08 & 89.05±0.01 \\
          & \textbf{Pubmed-LLM} & 71.75±0.14 & 73.54±0.27 & \textbf{90.38±0.32} & 79.99±0.11 & 78.91±0.14 & 76.87±0.08 & 78.41±0.18 & \textbf{90.27±0.34} & 82.19±0.07 & 81.93±0.13 \\
          & \textbf{Ogbn-arxiv-LLM} & 86.79±0.37 & 89.97±0.12 & 85.69±0.08 & \textbf{96.50±0.03} & \underline{89.74±0.04} & 86.55±0.35 & 89.91±0.10 & 86.94±0.06 & \textbf{96.50±0.03} & \underline{89.97±0.04} \\
          & \textbf{Our} & \textbf{88.19±0.48} & \textbf{92.28±0.28} & \underline{88.27±0.24} & \underline{96.30±0.01} & \textbf{91.26±0.25} & \textbf{88.59±0.41} & \textbf{92.39±0.28} & \underline{89.15±0.21} & \underline{96.30±0.01} & \textbf{91.61±0.22} \\
    \bottomrule
    \end{tabular}%
    }
  \label{dif_LLM}%
\end{table*}%

\begin{table*}[htbp]
  \centering
  \caption{Results of our proposed attack method for link stealing attacks on various GNN target models.}
        \scalebox{0.83}{
    \begin{tabular}{cc|ccccc|ccccc}
\cmidrule{3-12}          &       & \textbf{Cora} & \textbf{Citeseer} & \textbf{Pubmed} & \textbf{Ogbn-arxiv} & \textbf{Mean} & \textbf{Cora} & \textbf{Citeseer} & \textbf{Pubmed} & \textbf{Ogbn-arxiv} & \textbf{Mean} \\
    \midrule
    \multirow{5}[2]{*}{\textbf{GAT}} & \textbf{Cora-LLM} & 89.40±0.07 & 92.14±0.08 & 76.00±0.03 & 89.87±0.01 & 86.85±0.00 & 89.19±0.06 & 92.08±0.09 & 80.50±0.03 & 88.91±0.01 & 87.67±0.00 \\
          & \textbf{Citeseer-LLM} & 86.69±0.20 & \textbf{93.34±0.01} & 80.74±0.11 & 91.80±0.16 & 88.14±0.02 & 87.12±0.16 & \textbf{93.45±0.02} & 83.58±0.08 & 91.68±0.01 & 88.96±0.01 \\
          & \textbf{Pubmed-LLM} & 83.28±0.11 & 90.64±0.01 & \textbf{93.09±0.04} & 91.06±0.02 & \underline{89.52±0.04} & 84.23±0.06 & 90.93±0.01 & \textbf{93.10±0.04} & 91.00±0.02 & \underline{89.81±0.02} \\
          & \textbf{Ogbn-arxiv-LLM} & \underline{89.97±0.02} & 91.78±0.14 & 72.69±0.02 & \textbf{96.77±0.01} & 87.80±0.02 & \underline{90.22±0.05} & 92.15±0.11 & 78.45±0.02 & \textbf{96.77±0.01} & 89.39±0.01 \\
          & \textbf{Our} & \textbf{90.33±0.00} & \underline{92.91±0.04} & \underline{92.65±0.02} & \underline{93.42±0.04} & \textbf{92.32±0.02} & \textbf{90.53±0.01} & \underline{93.09±0.04} & \underline{92.70±0.02} & \underline{93.15±0.04} & \textbf{92.36±0.02} \\
    \midrule
    \multirow{5}[2]{*}{\textbf{SAGE}} & \textbf{Cora-LLM} & \underline{89.62±0.05} & 91.98±0.11 & 76.19±0.04 & 88.36±0.08 & 86.53±0.01 & 89.39±0.06 & 91.90±0.11 & 80.61±0.04 & 86.99±0.11 & 87.22±0.01 \\
          & \textbf{Citeseer-LLM} & 85.58±0.14 & \underline{92.55±0.14} & 77.62±0.11 & 91.36±0.04 & 86.78±0.09 & 86.02±0.14 & \underline{92.65±0.14} & 81.54±0.08 & 90.88±0.05 & 87.77±0.08 \\
          & \textbf{Pubmed-LLM} & 83.01±0.04 & 90.43±0.04 & \textbf{92.92±0.03} & 90.76±0.01 & \underline{89.28±0.00} & 83.90±0.01 & 90.65±0.04 & \textbf{92.91±0.04} & 90.41±0.01 & \underline{89.47±0.01} \\
          & \textbf{Ogbn-arxiv-LLM} & 89.35±0.21 & 91.50±0.00 & 73.57±0.05 & \textbf{96.91±0.01} & 87.83±0.04 & \underline{89.70±0.20} & 91.92±0.01 & 78.97±0.03 & \textbf{96.91±0.01} & 89.37±0.04 \\
          & \textbf{Our} & \textbf{90.23±0.04} & \textbf{93.02±0.09} & \underline{92.26±0.03} & \underline{94.58±0.01} & \textbf{92.52±0.02} & \textbf{90.40±0.03} & \textbf{93.21±0.08} & \underline{92.37±0.04} & \underline{94.48±0.01} & \textbf{92.61±0.02} \\
    \bottomrule
    \end{tabular}%
    }
  \label{dif_GNN}%
\end{table*}%

\subsubsection{Effect of Varying Maximum Magnitude $ \epsilon $}
Similarly, we explored the impact of varying maximum magnitudes $ \epsilon $ on model merging performance. The experimental results are presented in Fig. \ref{Magnitude} (b).

As shown in the figure, incorporating maximum magnitudes can enhance model performance in merging. However, the magnitude should not be set too high. When $ p = 0.5 $ and $ \epsilon = 0.1 $, the model merging achieves optimal results in link stealing attacks.

\subsection{Performance Analysis of the Proposed Method with Different Model Architectures}
Here, we explore the generality of our proposed method. First, we examine whether our method can be effectively applied across various LLM model architectures. Next, we investigate its effectiveness when performing attacks on different GNN architectures. The following section provides a detailed analysis of the experimental results.

\subsubsection{Attack Using Different LLM Architectures}
In the previous experiments, we adopted Vicuna-7B as the LLM architecture for conducting link stealing attacks. In this section, we further explore the use of other large language models, such as Vicuna-13B and LongChat, to examine the generality of our proposed method. The experimental results are presented in Table \ref{dif_LLM}. As shown in the table, even when employing different architectures of large language models, the proposed link stealing attack method effectively combines the advantages of each attacker's LLM-based attack model and achieves optimal average performance. A minimum mean accuracy of $91.26\%$ and an F1 score of $91.61\%$ are achieved across four datasets. Notably, when comparing attack performances across various datasets, our attack method sometimes achieves better performance than an attack model trained and tested on the same dataset. This outcome highlights the strong generalizability of our method, demonstrating its applicability across different LLM architectures.

\subsubsection{Attack on Different GNN Architectures}

Similarly, in addition to conducting link stealing attacks on the GCN target model in the previous experiments, we also carried out tests on other GNN target models, such as GAT and SAGE. The experimental results are shown in Table \ref{dif_GNN}. As indicated in the table, our attack method achieves an average accuracy and F1 score exceeding $92\%$ across models. On each dataset, our method consistently ranks in the top two for performance. These experimental results demonstrate that our method is applicable across different GNN target models and exhibits strong generalizability.

\subsection{Performance Analysis on Out-of-Domain Data}

\begin{figure}[tbp]
  \centering
  \subfigure[Accuracy]{
  \begin{minipage}[b]{0.23\textwidth}
    \includegraphics[width=\textwidth]{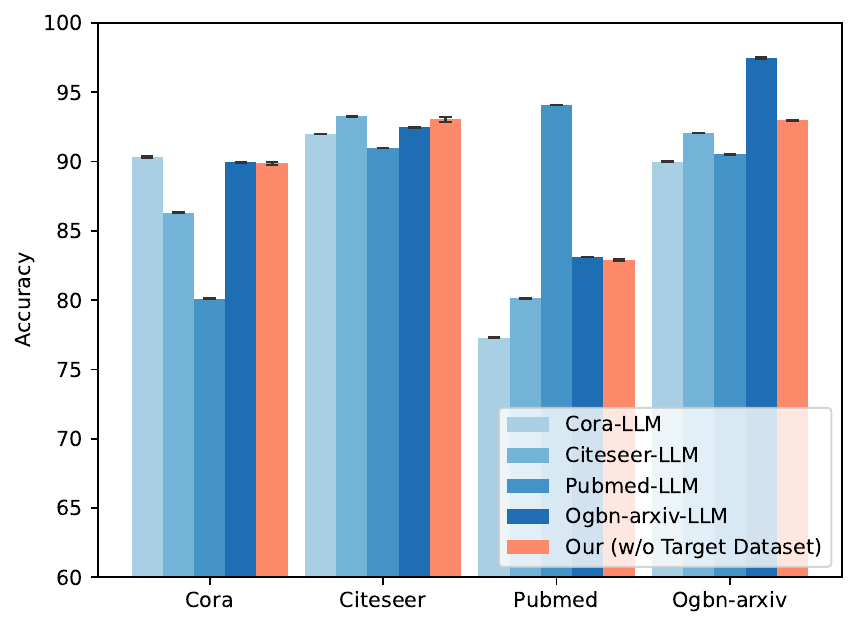}
  \end{minipage}
  }
  \subfigure[F1]{
  \begin{minipage}[b]{0.23\textwidth}
    \includegraphics[width=\textwidth]{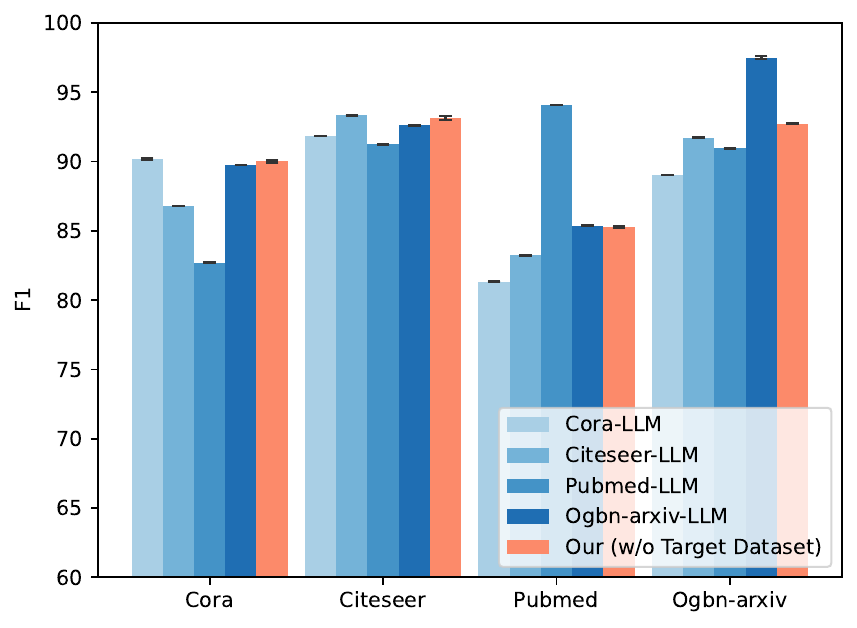}
  \end{minipage}
  }

  \caption{Comparison of our method using three attack models for merging and attacking out-of-domain data. For the attack on Cora, we perform model merging using the attack models trained on Citeseer, Pubmed, and Ogbn-arxiv. For the attack on Citeseer, we perform model merging using the attack models trained on Cora, Pubmed, and Ogbn-arxiv. For the attack on Pubmed, we perform model merging using the attack models trained on Cora, Citeseer, and Ogbn-arxiv. For the attack on Ogbn-arxiv, we perform model merging using the attack models trained on Cora, Citeseer, and Pubmed.}
  \label{three_merge}
\end{figure}

In this section, we examine the performance of the merging method on datasets unknown to the attackers, referred to as out-of-domain data. Additionally, we also conduct experiments on Cora, Citeseer, Pubmed, and Ogbn-arxiv to evaluate the generality of our method on out-of-domain data. To assess its effectiveness, we merge the attack models trained on three out of the four datasets and then use the resulting merged model to attack the remaining dataset. For example, we merge the attack models individually trained by attackers on the Citeseer, Pubmed, and Ogbn-arxiv datasets and evaluate the merged model's performance on the Cora dataset, which is considered out-of-domain since the attackers have no access to it. The experimental results are shown in Fig. \ref{three_merge}. In the figure, \emph{Our (w/o Target Dataset)} represents the model derived from merging the three attack models.

The figure demonstrates the effectiveness of our method on out-of-domain data. As shown, the merged model, derived from combining the three attack models, retains the best attack performance on out-of-domain data among the individual attack models. Specifically, when the Cora dataset is used as the out-of-domain dataset, the merged model achieves attack performance comparable to that of the attack model trained on Ogbn-arxiv for attacking Cora. Similarly, when the Citeseer and Pubmed datasets are treated as out-of-domain datasets, the merged model sustains the highest attack performance observed with the attack model trained on Ogbn-arxiv. Finally, when Ogbn-arxiv is used as the out-of-domain dataset, the merged model maintains the best attack performance of the attack model trained on Citeseer for targeting Ogbn-arxiv.

These findings suggest that our merging method effectively retains the highest performance on out-of-domain data among the models included in the merging process. This capability poses a significant threat to unknown datasets and aligns well with real-world attack scenarios.

\section{Related work}
\subsection{Privacy Attacks on Graph Neural Networks}
Graph neural networks (GNNs) \cite{DBLP:journals/air/GuanZZC24} contain a wealth of private information, which has led many researchers to focus on privacy attacks targeting these models~\cite{10052767,TCE2024}. In such attacks, adversaries can extract model details and sensitive data by querying deployed GNNs. Studies like \cite{DBLP:conf/asiaccs/WuYPY22,DBLP:conf/sp/ShenHH022,GUAN2024112144} have explored model extraction attacks on GNNs, where attackers utilize the target model’s posterior probabilities or intermediate-layer embeddings to capture its information and construct a surrogate model that closely replicates the original model’s functionality and performance.

In addition, \cite{DBLP:journals/corr/abs-2102-05429, DBLP:conf/icdm/WuYPY21} have investigated membership inference attacks on GNNs, focusing on determining whether specific data instances were included in the target model’s training set. These works build a shadow model that replicates the target GNN, allowing attackers to generate a shadow training dataset. By comparing the shadow model’s responses to those of the target model on similar data, attackers can infer if particular graph data was part of the target GNN’s training set. Zhang et al. \cite{zhang2022inference} introduced a novel variation of membership inference attacks that, instead of identifying individual data points in the target model’s training set, aims to determine whether a specific subgraph is embedded within a larger graph structure.

In attribute inference attacks on GNNs, attackers seek to infer specific or statistical information about the training graph data based on the target model’s responses. Wang et al. \cite{DBLP:conf/ccs/WangW22} conducted a comprehensive study on Group Property Inference Attacks (GPIA) within graph neural networks, focusing on two types of attributes: node group attributes, representing collective information of specific node groups, and link group attributes, representing collective information of specific link groups. Zhang et al. \cite{zhang2022inference} also investigated attribute inference attacks through graph embeddings, aiming to deduce fundamental attributes of the target graph, such as the number of nodes, number of edges, and graph density.

\textbf{Link Stealing Attacks: }A link stealing attack is a specific type of attribute inference attack where attackers extract information on node links in the training graph dataset by analyzing the target model’s responses. He et al. \cite{DBLP:conf/uss/HeJ0G021} first introduced the concept of link stealing attacks and examined their effectiveness under varying levels of attacker knowledge. In \cite{DBLP:conf/uss/HeJ0G021, DBLP:conf/icml/ZhangWWYXPY23,DBLP:journals/popets/WuHBHBGZ24}, researchers determined the presence of links between nodes by assessing the similarity of posterior probabilities obtained from the target model. Similarly, in \cite{DBLP:journals/corr/abs-2307-13548}, link inference was achieved by perturbing the graph data and comparing posterior probability changes before and after the perturbation.

However, previous studies have not addressed how to conduct cross-dataset link stealing attacks, nor have they explored methods for enabling multiple attackers to jointly execute such attacks. These issues remain open areas for further research.

\subsection{Large Language Models}
Large Language Models (LLMs) \cite{DBLP:conf/acl/LiYBZLSLSYWLXBF24} have transformed natural language processing, delivering outstanding performance in both academic and industrial contexts. LLMs, including models like GPT  \cite{DBLP:journals/mima/FloridiC20}, Vicuna \cite{DBLP:conf/nips/ZhengC00WZL0LXZ23} and LongChat \cite{longchat2023}, often contain hundreds of millions to billions of parameters, enabling them to capture intricate language patterns and contextual nuances. Given their superior capabilities, extensive research on LLMs has emerged across various fields.

Zeng et al. \cite{DBLP:conf/iclr/ZengLDWL0YXZXTM23} developed an open-source bilingual large language model that excels in both English and Chinese, leveraging specialized training strategies focused on selection, efficiency, and stability. Lin et al. \cite{DBLP:conf/eacl/LinWZLW24} explored the application of LLMs in media bias detection, addressing limitations of previous methods that relied on specific models and datasets, which restricted adaptability and performance on out-of-domain data. Liu et al. \cite{DBLP:conf/wsdm/LiuCS024} investigated the integration of LLMs into recommendation systems, enhancing content-based recommendation by incorporating both open-source and closed-source LLMs into the inference process.

In recent years, researchers have explored the integration of LLMs with GNNs to harness the complementary strengths of both architectures. He et al. \cite{DBLP:journals/corr/abs-2305-19523} leverage semantic knowledge from LLMs to improve the initial node embeddings in GNNs, enhancing performance in downstream tasks. Ye et al. \cite{DBLP:journals/corr/abs-2308-07134} replace the GNN predictor with an LLM, increasing the effectiveness of graph-based tasks by representing graph structures as natural language and using instruction prompts to utilize the expressive qualities of language. Liu et al. \cite{DBLP:journals/natmi/LiuNWLQLTXA23} align GNNs and LLMs within a shared vector space, incorporating textual knowledge into the graph, which enhances reasoning capabilities and overall model performance. Guo et al. \cite{guo2024graphedit} proposed GraphEdit, an approach that employs LLMs to capture complex relationships within graph-structured data, addressing the limitations of explicit graph structure information and enabling LLMs to learn graph structures effectively.

However, prior research has not explored the use of LLMs for privacy attacks on GNNs. In this paper, we introduce LLMs for link stealing attacks and design a specialized prompt to fine-tune the LLM, optimizing it specifically for link stealing tasks.

\subsection{Model Merging}
Model merging techniques combine multiple task-specific models into a unified, versatile model without necessitating data exchange or additional training \cite{DBLP:journals/corr/abs-2407-06089}. Yadav et al. \cite{DBLP:conf/nips/YadavTCRB23} tackled information loss in traditional merging methods, focusing on two main interference sources: redundant parameter values and conflicting parameter signs across models. Yu et al. \cite{DBLP:conf/icml/Yu0Y0L24} introduced DARE, a model merging approach that addresses parameter redundancy, especially within the highly redundant SFT delta parameters of LLMs. They proposed an efficient solution to reduce these delta parameters significantly without requiring data exchange, retraining, or even a GPU. Building on these advancements, Deep et al. \cite{DBLP:journals/corr/abs-2406-11617} introduced a magnitude-based method in model merging, where redundant parameter values are pruned, and the remaining parameters are rescaled, leading to enhanced model merging effectiveness.

However, these methods often overlook the relative strengths of each model, allowing poorly performing models to negatively impact the merged model's performance. In this paper, we propose a novel merging method to address this limitation. By assigning weights to each model based on its performance, our approach grants higher weights to better-performing models during the merging process, thus enhancing the final model's effectiveness.

\section{Conclusion}
In this paper, we propose a novel link stealing attack method that combines the knowledge of multiple attackers to perform cross-dataset link stealing attacks against graph neural networks. We introduce large language models to carry out cross-dataset attacks, addressing the challenge of varying data lengths in cross-dataset scenarios. To create a universal model by integrating knowledge from multiple attackers, we propose a novel model merging method. This approach merges attack models trained by individual attackers through three key steps: model dropping, parameter selection, and model merging. These steps allow attackers to effectively leverage the strengths of each attack model, enabling the merged model to deliver strong performance on in-domain data within the attackers' knowledge and achieve notable results on out-of-domain data previously unknown to the attackers. We provide theoretical proofs for the effectiveness of the proposed method and validate its performance through extensive experiments. We show that the proposed approach not only delivers excellent performance on in-domain data but also achieves effective attacks on out-of-domain datasets, aligning with real-world requirements. 

\bibliographystyle{IEEEtran}
\bibliography{bib}

\vfill

\end{document}